\documentclass[11pt]{article}
\usepackage[a4paper,margin=1in]{geometry}
\usepackage{url}

\usepackage{graphicx}
\usepackage{textcomp}
\usepackage{rotating}
\usepackage{xcolor}
\usepackage{booktabs}
\usepackage[table]{xcolor}  
\usepackage{colortbl}   
\usepackage{color}
\usepackage[most]{tcolorbox}
\usepackage{subcaption}
\usepackage{csquotes}
\usepackage{multirow}
\usepackage[normalem]{ulem}
\usepackage{todonotes}
\usepackage{lipsum}
\usepackage{float}
\usepackage{subfloat}
\usepackage{pgfplots, pgfplotstable}
\usepackage{tikz}
\pgfplotsset{compat=1.16}
\usepackage{pgfplotstable}
\usepackage{pgf-pie}
\usepackage{ragged2e}
\usepackage{adjustbox} 
\usepackage{ifthen}
\usepackage{soul}
\usepackage{array}
\usepackage{breakurl}
\usepackage{wasysym}

\usepackage[normalem]{ulem}

\usepackage{balance}
\usepackage{tikz}
\usepackage{pgf-pie}

\usepackage{pgfplots}
\usepackage{pgfplotstable}
\pgfplotsset{compat=1.18}
\usetikzlibrary{patterns} 

\tcbuselibrary{listings}
\usepackage{listings}
\usepackage{algorithm}       
\usepackage{algpseudocode}   

\usepackage{amsmath,amssymb} 

\usepackage{etoolbox}
\newtoggle{showpct}
\makeatletter
\patchcmd{\pgfpie@slice}%
{\pgfpie@scalefont{#3}\pgfpie@numbertext{#3}}%
{\iftoggle{showpct}{\pgfpie@scalefont{#3}\pgfpie@numbertext{#3}}{}}%
{}{}
\makeatother

\newtcolorbox[]{promptbox}[1][] { reset, #1}

\newtcolorbox{findingbox}{
  enhanced,
  colback=gray!10,                       
  colframe=gray!20,                      
  borderline west={3pt}{0pt}{gray!60},   
  sharp corners,
  boxrule=0pt,                           
  left=6pt,                              
  right=4pt,                             
  top=2pt,                               
  bottom=2pt,                            
  boxsep=0pt,                            
  before skip=6pt,
  after skip=6pt
}

\definecolor{codegreen}{rgb}{0,0.4,0}
\definecolor{codegray}{rgb}{0.5,0.5,0.5}
\definecolor{codepurple}{rgb}{0.58,0,0.82}
\definecolor{backcolour}{rgb}{1,1,0.97}
\definecolor{whitecolour}{rgb}{1,1,1}

\lstdefinestyle{mystyle}{
    backgroundcolor=\color{backcolour},   
    commentstyle=\color{codegreen},
    keywordstyle=\color{magenta},
    numberstyle=\tiny\color{codegray},
    stringstyle=\color{codepurple},
    basicstyle=\ttfamily\footnotesize,
    breakatwhitespace=false,         
    breaklines=true,                 
    captionpos=b,                    
    keepspaces=true,                 
    numbers=left,                    
    numbersep=5pt,                  
    showspaces=false,                
    showstringspaces=false,
    showtabs=false,                  
    tabsize=2
}

\usepackage[T1]{fontenc}

\usepackage{tcolorbox}

\newboolean{showcomments}
\setboolean{showcomments}{true} 

\ifthenelse{\boolean{showcomments}}
{
  \newcommand{\nbc}[3]{
    \colorbox{#3}{\bfseries\sffamily\scriptsize\textcolor{white}{#1}}
    {\textcolor{#3}{\sf\small$\blacktriangleright$\textit{#2}$\blacktriangleleft$}}
  }
}
{
  \newcommand{\nbc}[3]{}
}

\useunder{\uline}{\ul}{}
\def\BibTeX{{\rm B\kern-.05em{\sc i\kern-.025em b}\kern-.08em
    T\kern-.1667em\lower.7ex\hbox{E}\kern-.125emX}}

\AtBeginDocument{%
  \providecommand\BibTeX{{%
    Bib\TeX}}}

\title{Human-Aligned Enhancement of Programming Answers with LLMs Guided by User Feedback}

\author{
Suborno Deb Bappon \\
University of Saskatchewan, Canada \\
\texttt{subornodebbappon20@gmail.com}
\and
Saikat Mondal \\
University of Saskatchewan, Canada \\
\texttt{saikat.mondal@usask.ca}
\and
Chanchal K. Roy \\
University of Saskatchewan, Canada \\
\texttt{chanchal.roy@usask.ca}
\and
Kevin Schneider \\
University of Saskatchewan, Canada \\
\texttt{kevin.schneider@usask.ca}
}

\date{}
\begin{document}

\maketitle

\begin{abstract}
Large Language Models (LLMs) are increasingly used to assist software developers in diverse programming tasks (e.g., code generation, documentation, optimization). However, their potential to improve existing programming answers as human experts do remains underexplored. On Technical Q\&A platforms like Stack Overflow (SO), human contributors revise answers in response to user feedback that point out errors, inefficiencies, or missing explanations. However, around one-third of such feedback often remains unaddressed due to effort, expertise, time, or visibility constraints, which leaves many answers outdated or incomplete.
In this study, we investigate whether LLMs can perform human-aligned answer enhancement by interpreting and integrating comment feedback. Our contributions are fourfold. 
(1) We build \textit{ReSOlve}, a benchmark of 790 SO answers and their comment threads, 
 annotated for improvement-related and general feedback. 
(2) We evaluate four state-of-the-art LLMs (e.g., DeepSeek-V3.1, GPT-5, Gemini 2.5 Flash, and LLaMA-4 Maverick) on their ability to identify actionable (i.e., improvement-related) concerns, where DeepSeek achieves the most balanced precision–recall trade-off. 
(3) We introduce \textit{AUTOCOMBAT}, an LLM-powered tool that enhances programming answers by jointly leveraging one/multiple user comments and question context. Evaluation against human-revised references shows that \textit{AUTOCOMBAT} can perform near-human improvements while preserving the original intent, and significantly outperforms the baseline model.
(4) A user study with 58 practitioners further confirms the utility of the tool. In total, 84.5\% of participants reported that they would adopt or recommend AUTOCOMBAT. Participants attributed this to reduced manual effort, better coverage of issues raised in comments, and clearer explanations that more accurately incorporate user feedback.
Overall, \textit{AUTOCOMBAT} enables scalable integration of community feedback into SO answers, enhancing reliability, research utility, and platform trust. \\


\end{abstract}

\section{Introduction}
\label{sec:introduction}

Large Language Models (LLMs) have rapidly become integral to modern software engineering (SE) practices, assisting developers in code generation, debugging, documentation, and performance optimization \cite{hou2024large, zhang2023survey, fan2023large, ross2023programmer}. LLMs power tools such as GitHub Copilot and ChatGPT, which can generate useful code suggestions and explanations in real time \cite{chen2021evaluating, bubeck2023sparks, Fried2022InCoderAG, svyatkovskiy2020intellicode, vaithilingam2022expectation, mozannar2024reading}. After generating an initial output, developers often refine it through follow-up prompts or further interactions with the model. 
These refinements occur in an isolated and self-directed manner, where the same user provides the feedback, defines the goal, and decides when the result is \textit{good enough}.
However, such refinement capacities have not been systematically evaluated in wild, complex, and community-driven forums such as Stack Overflow (SO). 
On SO, developers submit initial answers to questions that receive diverse feedback identifying outdated libraries, inefficient logic, or potential security flaws in code 
\cite{Jallow2024Measuring, ragkhitwetsagul2019toxic}. Textual explanations that accompany the code also affect usability but are often incomplete or ambiguous, which limits their educational and practical value \cite{Treude2017Understanding, Zhang2022A, Zerouali2021Identifying, Ponzanelli2014Improving, baltes2018sotorrent}. Given the rapid progress of LLMs in understanding and improving both code and natural language, it is a timely attempt to examine the extent to which they can automate such feedback-driven improvements within real software development communities.

Comments posted against answers on SO often contain valuable observations that identify coding errors, outdated practices, alternative solutions, or unclear explanations \cite{ragkhitwetsagul2019toxic, Zhang2019AnES, Nie2017DataDrivenAS, Gao2020TechnicalQS, soni2019analyzing, Zhang2021ReadingAO}. However, prior research shows that nearly one-third of actionable feedback remains unaddressed \cite{sheikhaei2023study}, leaving many answers outdated or incomplete. 
Addressing these concerns manually is often impractical because it requires significant time, contextual understanding, and expertise, especially when feedback spans multiple aspects that are difficult to synthesize into a cohesive solution. 

\begin{figure*}[!htbp]
    \centering
    \includegraphics[width=0.8\textwidth]{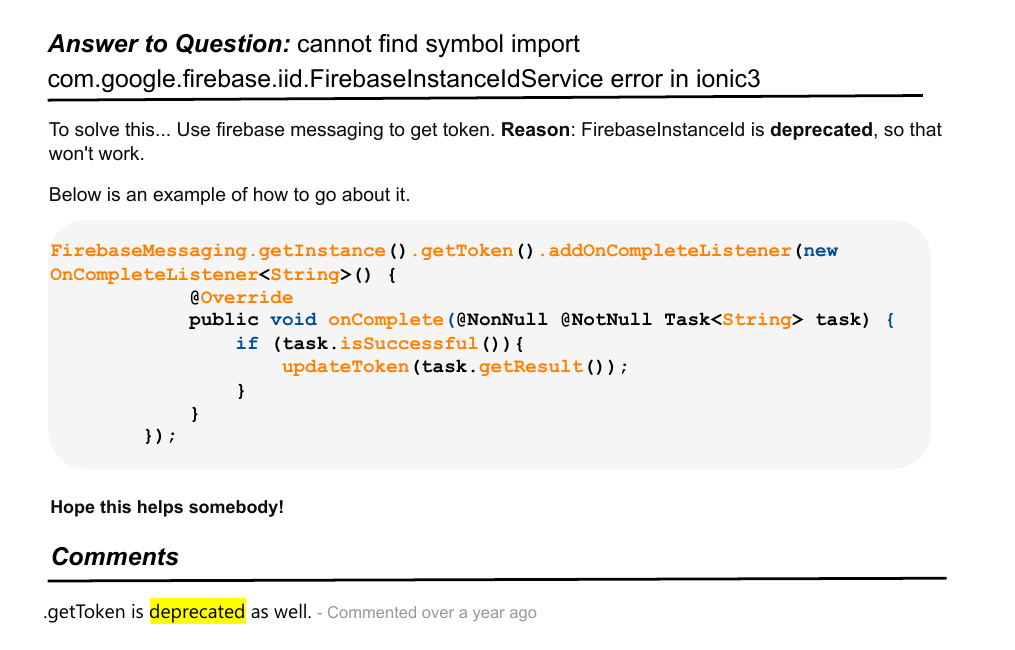} 
    \caption{Motivating example.}
    \label{fig:intro-mot}
\end{figure*}

For example, as illustrated in Fig. \ref{fig:intro-mot}, a comment on an SO answer \cite{StackOverflowFirebaseSymbolError} reported that \texttt{.getToken()} was deprecated, but the answer remained unrevised for years, motivating the need for automated methods to detect and address improvement-related comments.

In addition, SO's default interface only displays a few comments, causing important suggestions to go unnoticed \cite{Zhang2019Does}. As a result, developers who rely on these unedited or outdated answers may propagate bugs, inefficiencies, or security issues into their projects, which can negatively impact software quality and maintainability \cite{Gerasimou2018On, xie2008improving, Mai2024TowardsBA}. An automated approach is therefore essential to improve the accuracy, relevance, and long-term reliability of SO's programming knowledge base.

Existing studies leveraged LLMs to generate inline code comments for SO answers \cite{Bappon2024AUTOGENICSAG} or to generate answers from scratch \cite{xu2023we}, but enhancing answers based on developer feedback remains under-examined. A recent study introduced SOUP \cite{Mai2024TowardsBA}, a framework that updates SO answer code snippets using user feedback. For answers involving a single comment and a single code snippet, SOUP demonstrates promising performance in improving code snippets. However, SOUP has several practical limitations that affect its applicability in more realistic settings. It updates only code snippets while ignoring the textual explanations that are critical for understanding a solution. It cannot handle answers that contain multiple code snippets, which restricts its general applicability. It can address single-comment feedback and fails to combine insights from several comments that may overlap or conflict. It also disregards the question context during refinement, leading to syntactically correct revisions that might be semantically inconsistent with the original intent. 
Therefore, a more comprehensive approach is needed that can identify actionable concerns across diverse comment threads, synthesize multiple feedback signals, and refine both the code and its explanation while maintaining alignment with the corresponding question, as a human expert would.

In this study, we introduce \textit{AUTOCOMBAT} (AUTOmated COMment Based Answer enhancemenT), a tool that leverages LLMs to achieve human-aligned enhancement of SO answers by addressing user feedback extracted from comments and refining both code and explanations, while preserving relevance through integration of the corresponding question context.
We constructed a benchmark dataset of 790 SO answers, where we manually labeled their comments as improvement-related and addressed (\textit{IA}), improvement-related but not addressed (\textit{INA}), and general comments (\textit{GC}). For each answer, we collected the initial version (i.e., before any improvement-related comments labeled as \textit{IA} were addressed) and the revised version (i.e., after addressing the relevant improvement-related comments labeled as \textit{IA}).
We then employed state-of-the-art LLMs, including DeepSeek-V3.1 (Reasoner) \cite{DeepSeekV31}, GPT-5 \cite{GPT5}, Gemini 2.5 Flash \cite{Gemini25Flash}, and LLaMA-4 Maverick \cite{LLaMAMaverick} to identify improvement-related comments and refine initial answers while leveraging question context when required. 
The refined outputs were evaluated against human-revised ground truth (i.e., revised answers) using a comprehensive suite of syntactic and semantic metrics. In addition, we conducted a human evaluation to assess whether the refined answers preserved the original intent and behavior of the solutions. Finally, we surveyed with 58 professional developers to assess the acceptability and usefulness of \textit{AUTOCOMBAT} in practice. In particular, this study addressed three research questions and contributes new insights into feedback-aware, human-aligned answer refinement using LLMs.

\textbf{\textbf{RQ1:} To what extent can LLMs identify improvement-related concerns expressed in comments on programming answers?} Effectively identifying improvement concerns within comment threads is crucial for guiding answer refinements in the right direction. To assess this capability, we evaluate how well LLMs, such as DeepSeek-V3.1 (Reasoner), GPT-5, Gemini 2.5 Flash, and LLaMA-4 Maverick can detect user comments that point to potential improvements. The evaluation is conducted using six key metrics: accuracy, precision, recall, F1-score, specificity, and Matthews Correlation Coefficient (MCC).

\textbf{\textbf{RQ2:} To what extent can \textit{AUTOCOMBAT}, designed with LLM capabilities, achieve improvements comparable to human revisions in programming answers?}
Addressing user feedback while preserving alignment with the question context is essential to ensure that answer refinements remain relevant and contextually accurate.
To address this, we introduced \textit{AUTOCOMBAT}, a browser plugin that refines SO answers by integrating improvement-related concerns aligned with the corresponding question context. We evaluate the refined answers by comparing them against human-revised answers using a set of automated metrics--covering both syntactic measures (e.g., ROUGE \cite{Lin2004ROUGEAP}, BLEU \cite{Papineni2002BleuAM}, METEOR \cite{Banerjee2005METEORAA}, TER \cite{Snover2005ASO}, SacreBLEU \cite{post-2018-call}, chrF \cite{popovic2015chrf}, Jaccard Index \cite{real1996probabilistic}, Distinct-n \cite{li-etal-2016-diversity}, TF-IDF Cosine Similarity \cite{salton1988term}) and semantic measures (e.g., BLEURT \cite{Sellam2020BLEURT}, COMET \cite{Rei2020COMET}, BARTScore \cite{Yuan2021BARTScore}, BERTScore\_F1 \cite{Zhang2020BERTScore}, MoverScore \cite{Zhao2019MoverScore}). In addition to automated evaluations, we perform a manual human assessment of intent preservation to determine whether the refined answers maintained the original technical goal and solution behavior with respect to the corresponding question, as reflected in the human-revised reference answers. This evaluation captures subtle intent drift and behavioral misalignment that automated similarity metrics cannot reliably detect.

\textbf{\textbf{RQ3:}  How do users perceive the acceptability and usefulness of answers enhanced by \textit{AUTOCOMBAT} in practical settings?} Understanding how users perceive enhanced answers is essential to validate their practical value and acceptance in real-world settings. To assess this, we evaluated the real-world usefulness of \textit{AUTOCOMBAT} by surveying 58 developers, who rated the enhanced answers based on \textit{coverage}, \textit{clarity}, and \textit{relevance}.

This work advances a broader paradigm of human-LLM collaboration in software knowledge maintenance, where LLMs act not as autonomous code writers but as responsive partners that synthesize dispersed human feedback into sustained improvements. By operationalizing this paradigm on SO, \textit{AUTOCOMBAT} establishes a foundation for feedback-driven software intelligence that continuously aligns with community norms and technical evolution.

\smallskip
\noindent The replication package for this study will be made publicly available.



\section{Methodology}
\label{sec:methodology}

\begin{figure*}[!htbp!]
    \centering
    \includegraphics[width=\textwidth]{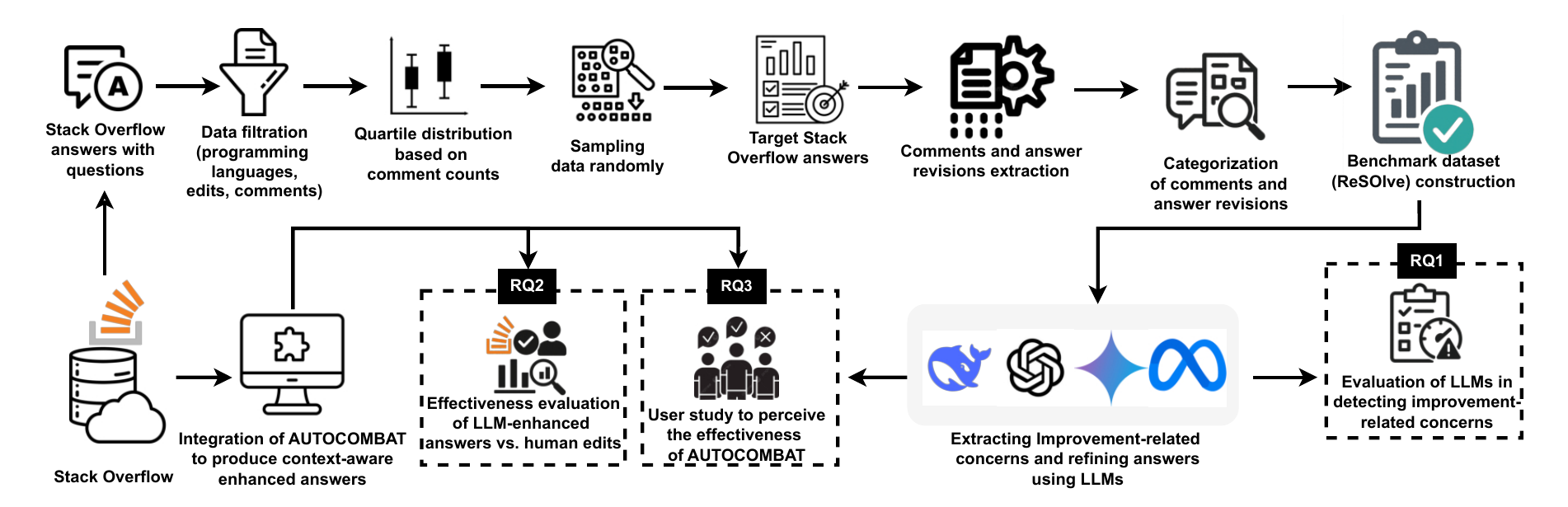} 
    \caption{Study methodology.}
    \label{fig:methodology}
\end{figure*}

Fig. \ref{fig:methodology} shows the schematic diagram of our study methodology. The following sections describe each step of the methodology in detail.

\subsection{Dataset Preparation}

\textbf{Data Collection \& Filtration.} Table \ref{tab:dataset_prep} summarizes the dataset used in this study. We collected 246,541 SO answers and their corresponding questions using the StackExchange API \cite{stackoverflow2024}, covering posts up to March 2025.
In particular, we collected answers to questions related to four popular programming languages, namely Python, Java, C\#, and JavaScript, which represent both static and dynamic paradigms. 
We target answers that were enhanced in response to concerns raised in their comments, and thus excluded answers that had no comments. Later, we compared LLM-enhanced answers based on comment-driven concerns with human-revised answers that addressed the same concerns (Section \ref{subsec:human-comparable-ans-improvement}). We thus only considered those answers that had at least one edit. After applying these filters, our final dataset consisted of 27,552 answers. Although the StackExchange API provides only the latest versions of answers, to analyze the other versions, we saved the webpages containing revision histories and parsed them using appropriate HTML tags (e.g., \texttt{div.js-revisions}) to extract all versions of each answer.

\begin{table}[!htbp]
\centering
\caption{Summary of our dataset (\textit{Q1--Q4}: Quartiles based on comment counts).}
\label{tab:dataset_prep}
\resizebox{\textwidth}{!}{
\begin{tabular}{lcccc|c|cccc|cccc}
\toprule
& \multicolumn{4}{c|}{\textbf{Total Answers}} 
& \textbf{Answers with Edits} 
& \multicolumn{4}{c|}{\textbf{Quartile Analysis (Comment Count Ranges)}} 
& \multicolumn{4}{c}{\textbf{Sampled Answers}} \\ 
\cmidrule(lr){2-5} \cmidrule(lr){7-10} \cmidrule(lr){11-14}
& \textbf{Python} & \textbf{Java} & \textbf{C\#} & \textbf{JavaScript} 
& \textbf{Count} 
& \textbf{Q1 (1)} & \textbf{Q2 (2--3)} & \textbf{Q3 (4--5)} & \textbf{Q4 (6--13)} 
& \textbf{Q1} & \textbf{Q2} & \textbf{Q3} & \textbf{Q4} \\ 
\midrule
\textbf{Count} 
& 70,371 & 55,693 & 51,846 & 68,631 
& 27,552 
& 7,344 & 9,022 & 4,713 & 6,473 
& 250 & 250 & 250 & 250 \\ 
\bottomrule
\end{tabular}
}
\end{table}

\noindent\textbf{Quartile Distribution \& Sampling.} To capture different levels of answer refinement, from few concerns to many, we counted the comments for each of the 27,552 SO answers. The distribution of comment counts was then divided into quartiles: Q1 (1), Q2 (2–3), Q3 (4–5), and Q4 (6–13), as shown in Table \ref{tab:dataset_prep}. Finally, we randomly selected 1000 answers (250 from each quartile), which is statistically significant at the 95\% confidence level with a 5\% error margin \cite{Boslaugh2008StatisticsIA, Mondal2024CanWI}.

\noindent\textbf{Qualitative Analysis.} 
We then conducted a qualitative analysis of the associated comments and revisions of these 1000 sampled answers. Two annotators, the first author (over six years of professional software development experience) and a research intern (four years of experience), participated in this manual analysis. Prior to annotation, both participated in calibration sessions with 20 example cases to ensure consistency in the following two subtasks: (1) carefully analyzed and categorized the comment threads into \textit{IA}, \textit{INA}, or \textit{GC}, and (2) identifying the corresponding initial and revised answer revisions. These examples were not included in our selected 1000 samples. 

\begin{algorithm}[!htbp]
\caption{Curation of the \textit{ReSOlve} Benchmark Dataset}
\label{alg:resolve_curation}
\begin{algorithmic}[1]

\Require Version history $V=\{v_1,\dots,v_n\}$, comments $C=\{c_1,\dots,c_m\}$
\Ensure $(v_{\text{init}}, v_{\text{final}})$ and $C_{\text{rel}}$
\Statex \textit{Assume $V$ and $C$ are ordered by timestamp.}

\State Partition $C$ into $C_{\text{IA}}$ (improvement and addressed), $C_{\text{INA}}$ (improvement and not addressed), and $C_{\text{GC}}$ (generic).
\If{$C_{\text{IA}}=\emptyset$}
    \State $v_{\text{init}} \gets \text{latest}(V)$; \ $v_{\text{final}} \gets v_{\text{init}}$; \ $C_{\text{rel}} \gets C_{\text{GC}}$
    \State \Return $(v_{\text{init}}, v_{\text{final}}, C_{\text{rel}})$
\EndIf

\State $c^* \gets \text{earliest}(C_{\text{IA}})$; \ $t^* \gets t(c^*)$
\State $v_{\text{init}} \gets \max\{\, v \in V \mid t(v) < t^* \,\}$

\State $V' \gets \{\, v \in V \mid \exists c \in C_{\text{IA}} \text{ such that } v \text{ incorporates the concern in } c \,\}$
\State $v_{\text{final}} \gets 
\begin{cases}
\text{latest}(V'), & \text{if } V' \neq \emptyset \\
v_{\text{init}}, & \text{otherwise}
\end{cases}$

\State $C_{\text{rel}} \gets (C_{\text{IA}} \cup C_{\text{GC}}) \setminus C_{\text{INA}}$
\State \Return $(v_{\text{init}}, v_{\text{final}}, C_{\text{rel}})$

\end{algorithmic}
\end{algorithm}

Annotators first conducted manual labeling of the comments associated with 1,000 sampled answers. Inter-rater agreement was measured using Cohen’s $\kappa$ \cite{Cohen1960ACO, Cohen1968WeightedKN}, yielding $\kappa$ = 0.92 (i.e., almost perfect agreement) for comment classification. In particular, the annotators disagreed on 84 instances, deciding \textit{IA} or \textit{INA}. However, all disagreements were resolved through discussion to reach a consensus. Next, the annotators independently identified the corresponding initial version (the answer just before the first improvement-related concern was addressed) and revised version (after addressing all improvement-related concerns) of each answer. In this case, the inter-rater agreement was again perfect ($\kappa$ = 0.91). Fifty-two disagreements were mostly due to deciding whether an edit was triggered by a specific comment. 

We retained the \textit{IA} category since they were addressed, and the \textit{GC} category, as they provide context for distinguishing useful comments from general discussion. However, the \textit{INA} comments were excluded because they were not reflected in edits. Including them could make the ground truth inconsistent, as LLMs might respond to feedback that humans did not address. In this way, we constructed our benchmark dataset, \textbf{\textit{Refinement of SO Answers through Community Feedback (ReSOlve)}}, which includes 790 instances where comments are labeled as \textit{IA} and \textit{GC}, with corresponding initial and revised answer versions. These are distributed across quartiles as: 122 in Q1, 206 in Q2, 224 in Q3, and 238 in Q4. The lower count in Q1 occurs because it contains one comment and many of them were \textit{INA.}
In total, the benchmark creation required about 150 person-hours. Algorithm \ref{alg:resolve_curation} illustrates the complete dataset curation procedure.

\subsection{Identifying Improvement-Related Concerns in Comments (RQ1)}
\label{subsec:identifying-improvement-concerns}

We examined how effectively LLMs can identify useful comments that help answers improvement. 
Formally, given a set of comments $C = \{c_1, c_2, \dots, c_n\}$ associated with an answer, 
the task is to partition $C$ into a set of improvement-related comments 
$C_{IA} = \{IA_1, IA_2, \dots, IA_k\}$ and a set of general comments 
$C_{GC} = \{GC_1, GC_2, \dots, GC_m\}$, such that $C_{IA} \cup C_{GC} = C$ and 
$C_{IA} \cap C_{GC} = \emptyset$. 
Each LLM induces a mapping $f_{\text{LLM}} : C \rightarrow \{\textit{IA}, \textit{GC}\}$  that assigns a label to each comment.

To evaluate this capability, we first removed the manually assigned labels (i.e., \textit{IA} and \textit{GC}) from all comments of our benchmark, \textit{ReSOlve}. We then used four state-of-the-art LLMs, namely DeepSeek, GPT, Gemini, and LLaMA, to identify improvement-related concerns within these comments. We designed an optimal prompt through several trials and adjustments \cite{sabbatella2024prompt, sahoo2024systematic}, as shown in Table~\ref{tab:answer_refinement}, and used it to guide all LLMs in identifying improvement-related concerns. The output from the LLMs was a list of potential improvement-related concerns, formatted in a JSON file. The concerns were either extracted as substrings from the comments, paraphrased, or supplemented with additional context by the models.
%

%
Subsequently, we manually reviewed the output of each model by comparing the identified concerns with the manually classified categories (\textit{IA} and \textit{GC}). Formally, let $y(c_i) \in \{\textit{IA}, \textit{GC}\}$ denote the gold-standard label of a comment $c_i \in C$, and let $\hat{y}_{\text{LLM}}(c_i) = f_{\text{LLM}}(c_i)$ denote the label predicted by an LLM. Model performance was evaluated by comparing $\hat{y}_{\text{LLM}}(c_i)$ against $y(c_i)$ for all comments.

Based on this comparison, we constructed a confusion matrix to assess how effectively the LLMs detected improvement-related concerns. In this matrix, \textit{True Positives (TP)} denote comments identified as \textit{IA} that are originally \textit{IA}, \textit{False Positives (FP)} denote comments identified as \textit{IA} that are originally \textit{GC}, \textit{True Negatives (TN)} denote comments identified as \textit{GC} that are originally \textit{GC}, and \textit{False Negatives (FN)} denote comments identified as \textit{GC} that are originally \textit{IA}. These quantities were computed as follows:

\[
\begin{aligned}
TP &= |\{c_i \mid \hat{y}_{\text{LLM}}(c_i)=\textit{IA} \wedge y(c_i)=\textit{IA}\}|, \\
FP &= |\{c_i \mid \hat{y}_{\text{LLM}}(c_i)=\textit{IA} \wedge y(c_i)=\textit{GC}\}|, \\
TN &= |\{c_i \mid \hat{y}_{\text{LLM}}(c_i)=\textit{GC} \wedge y(c_i)=\textit{GC}\}|, \\
FN &= |\{c_i \mid \hat{y}_{\text{LLM}}(c_i)=\textit{GC} \wedge y(c_i)=\textit{IA}\}|.
\end{aligned}
\]

Finally, we calculated standard classification metrics derived from this confusion matrix, including precision, recall, F1-score, accuracy, specificity, and MCC for each LLM.



\subsection{Assessing LLM's Human-Comparable Answer Improvements (RQ2)}
\label{subsec:human-comparable-ans-improvement}


\textbf{Generation of Improved Answers using LLMs.} To refine the initial versions of SO answers, we utilized the same four LLMs described earlier. Each answer was paired with its associated comments (\textit{IA} and \textit{GC}), which served as input for identifying improvement-related concerns that guided the refinement process. While comments provide actionable suggestions, the question context remains crucial for understanding the asker’s intent and technical constraints \cite{Bappon2024AUTOGENICSAG, Galappaththi2022DoesTA, prenner2024out}. To ensure well-informed revisions, each model was given the autonomy to decide whether to use the question context alongside comment feedback. The refinement was performed using a structured prompt, shown in Table~\ref{tab:answer_refinement}, which defined the editing policies, and response format.

Formally, let $q$ denote a SO question, $a^{\text{init}}$ the initial answer, and $C = \{c_1, c_2, \dots, c_n\}$ the associated comment set. Given the subset of improvement-related concerns $C_{IA} \subseteq C$, each LLM defines a refinement function
\[
r_{\text{LLM}} : (a^{\text{orig}}, C_{IA}, q) \rightarrow a^{\text{LLM}},
\]
where $a^{\text{LLM}}$ is the refined answer produced by the model. The question context $q$ is optionally incorporated by the model only when required to resolve the identified concerns, in accordance with the prompt policies.

\begin{table}[!htbp]
\caption{Answer refinement prompt.}
\label{tab:answer_refinement}
\centering
\begin{tabular}{p{0.96\textwidth}}
    \toprule
    \noindent You are a Stack Overflow Answer Refiner. \\ \\
    \textbf{Policy:} \\
    $\bullet$ Edit ONLY if comments contain actionable, improvement-related concerns \newline
    $\bullet$ Do NOT add your own corrections or ideas. Do not evaluate technical correctness \newline
    $\bullet$ Even if a suggested change appears incorrect, apply it if it is an actionable improvement request\newline
    $\bullet$ You MAY read/use the question ONLY if needed to resolve those concerns \newline
    $\bullet$ Keep edits minimal, targeted, and concise; preserve original answer structure; SO tone; fenced code when useful\newline
    $\bullet$ Do not fabricate APIs/behavior/version claims. If a detail is not in comments (or necessary question context), do not invent it \\ \\

    \textbf{Output JSON only:} \\
    \textit{\{ 
    "concerns": ["actionable concerns"], \newline
    "used\_question": true|false, \newline
    "change\_log": [{"concern": "...", "change": "..."}], \newline
    "improved\_answer": "final answer text" 
    \}} \\ \\
    
    \textbf{User Prompt Template:} \\
    \textit{Original Answer: \{answer\}; \newline 
    Comments (mix of actionable + generic; order preserved as provided): \{comments\}; \newline 
    Question (use ONLY if needed): \{question\}} \\ \\
    
    \textbf{Tasks:} \\
    1) Extract only actionable improvement concerns from comments (ignore thanks, jokes, meta, vague, generic remarks). \\
    2) Set used\_question=true if you needed the question to resolve concerns; else false. \\
    3) Produce a revised answer addressing ONLY those concerns; minimal edits; keep helpful structure. \\
    4) If no actionable concerns, return the original answer unchanged. \\
    5) If concerns conflict, follow the later one based on the given order. 
    \\ \bottomrule
    \end{tabular}
\end{table}

\textbf{Effectiveness Evaluation of LLMs in Enhancing SO Answer.} To evaluate the effectiveness of these LLM-enhanced answers, we compared them against the revised version of SO answers from \textit{ReSOlve} benchmark using two complementary strategies: (1) automated metric-based evaluation to assess syntactic and semantic similarity, and (2) manual evaluation to examine intent preservation.  

\begin{itemize}
    \item \textit{Automated Evaluation of LLM-refined Answers:} We performed an automated evaluation using a comprehensive suite of syntactic and semantic similarity metrics. In this evaluation, human-revised answers served as references, while LLM-enhanced answers were treated as predictions. For each SO answer, let $a^{\text{orig}}$ denote the initial answer, $a^{\text{LLM}}$ the LLM-enhanced answer, and $a^{\text{human}}$ the human-revised reference answer. The evaluation task was to assess the similarity between $a^{\text{LLM}}$ and $a^{\text{human}}$, treating the latter as ground truth.
    
    For syntactic similarity, we calculated ROUGE-(1 to 2) for phrase overlap, ROUGE-L for shared sequences, BLEU-(1 to 4) for n-gram accuracy, METEOR for flexible matching, TER score for edit count, SacreBLEU for consistency, chrF for character n-grams, Jaccard for set overlap, Dist-(1 to 2) for n-gram differences, and TF-IDF cosine for word importance. For semantic similarity, we employed BLEURT for contextual understanding, COMET for quality metrics, BARTScore for coherence, BERTScore\textunderscore F1 for token alignment, and MoverScore for semantic distance. 
    
    For each metric $m \in \mathcal{M}$, we computed a score $m(a^{\text{LLM}}, a^{\text{human}})$, and aggregated results across all evaluated answers to compare model performance. For TER, lower scores indicate better performance, whereas for all other metrics, higher scores indicate better performance.
    Based on the performance of the best-performing model, we introduced \textit{AUTOCOMBAT}.

    \item \textit{Manual Evaluation of Intent Preservation:} To complement the automated syntactic and semantic evaluations, we conducted a manual assessment of intent preservation to determine whether LLM-enhanced answers maintained the technical goal pursued by the human-revised answer with respect to the corresponding question. While automated metrics capture lexical overlap and semantic similarity, they are insufficient for reliably identifying intent drift, particularly in cases where refinements subtly alter problem scope, assumptions, or focus \cite{evtikhiev2023out, hu2022correlating}. We therefore incorporated a human-centered evaluation to explicitly assess alignment with the original intent.
    
    We (two authors of this paper) manually investigated all LLM-enhanced answers in our benchmark. The first annotator was the lead author, with over six years of professional software development experience. The second annotator was the second author, with sixteen years of programming experience and seven years of software engineering research experience. Both annotators were familiar with interpreting SO questions, answers, and associated comment threads.

    Before conducting the annotation, we discussed the notion of intent preservation in multiple interactive sessions. We then analyzed and labeled 100 randomly selected examples drawn from the evaluation set to calibrate our understanding. Intent was defined as the technical goal pursued by the answer in addressing the corresponding question, as reflected in the human-revised reference answer. Annotators were instructed to ignore differences in wording, formatting, or coding style, and to focus solely on whether the refined answer pursued the same goal under the same assumptions.

    For each answer instance, let $q$ denote the original question, $a^{\text{human}}$ the human-revised reference answer, and $a^{\text{LLM}}$ the LLM-enhanced answer. The intent preservation task is to assess whether $a^{\text{LLM}}$ preserves the same technical intent as $a^{\text{human}}$ with respect to $q$. Each annotator defines a labeling function
    \[
    g : (q, a^{\text{human}}, a^{\text{LLM}}) \rightarrow \{\textit{YES}, \textit{PARTIALLY\ YES}, \textit{NO}\},
    \]
    which assigns an intent-preservation label to each LLM-enhanced answer.
    
    For each LLM-enhanced answer, we carefully examined whether the refinement fully preserved the original intent, partially altered important constraints or scope, or changed the intent entirely. Based on this analysis, each answer was assigned one of three labels: \emph{YES}, \emph{PARTIALLY YES}, or \emph{NO}. Answers were labeled \emph{YES} if the original intent was fully preserved, \emph{PARTIALLY YES} if the general goal was maintained but important constraints or scope were altered, and \emph{NO} if the original intent was replaced, contradicted, or no longer recognizable.

    We then measured inter-annotator agreement using Cohen’s Kappa \cite{Cohen1960ACO}. The resulting $\kappa$ value was 0.95, indicating almost perfect agreement between the annotators. We resolved the remaining few disagreements through discussion. The high level of agreement suggests that the labeling criteria are clear and can be applied consistently without introducing individual bias \cite{Bappon2024AUTOGENICSAG}. Accordingly, the first author completed the labeling of the remaining 690 cases following the agreed-upon criteria. The entire annotation process required approximately 100 person-hours.

\end{itemize}

\begin{figure*}[!htbp]
    \centering
    \includegraphics[width=\textwidth]{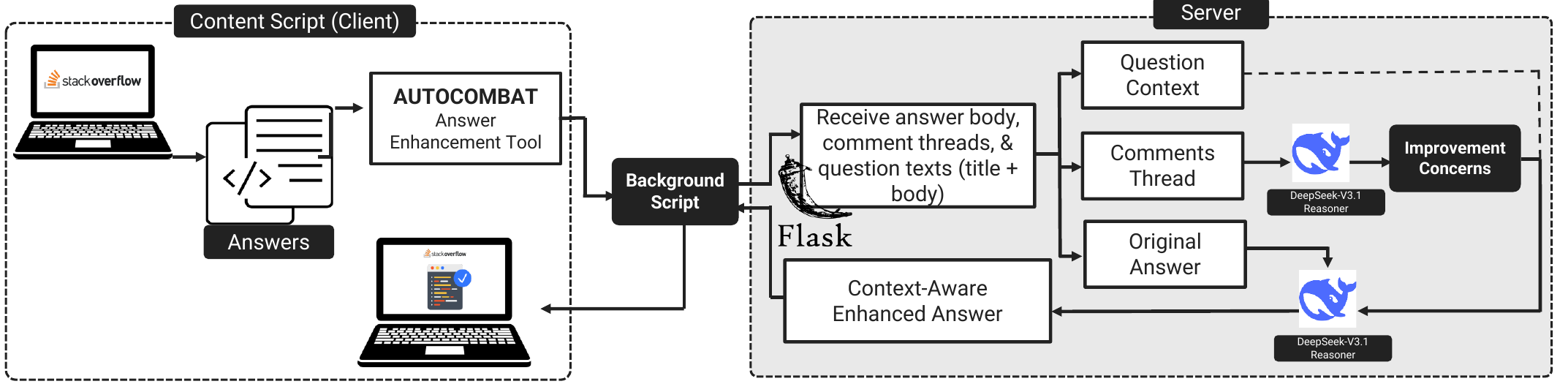} 
    \caption{An overview of \texttt{AUTOCOMBAT} system architecture.}
    \label{fig:arch_autocombat}
\end{figure*}

\textbf{System Architecture of \textit{AUTOCOMBAT}.} 
\textit{AUTOCOMBAT} is a browser plugin built on the best-performing \textbf{DeepSeek-V3.1} model. It enhances SO answers by addressing comment-driven feedback and autonomously incorporating relevant question context when appropriate. Fig. \ref{fig:arch_autocombat} shows the overview of its architecture.

The \textit{front-end} of the plugin is implemented as a Chrome browser extension that activates on SO question pages. Upon loading, the extension’s content script scans each answer for associated comments. If comments are detected, it dynamically inserts an \textit{``AUTOCOMBAT''} button beneath the corresponding answer. When a user clicks this button, the content script collects the answer text, its associated comments, and the corresponding question description from the page’s DOM. This data is then organized into a structured JSON object and transmitted to the backend server via an HTTP \textit{POST} request, routed through a background script.

The \textit{backend server}, built with Flask, processes incoming requests through a series of steps. Upon receiving the answer, its associated comments, and corresponding question, the server constructs a unified refinement prompt following predefined editing policies described in Table \ref{tab:answer_refinement}. This prompt is sent to the \texttt{deepseek\_reasoner} model through its OpenAI-compatible API~\cite{api_deepseek}, secured by an authentication key. The model operates at a temperature of zero to ensure deterministic and consistent outputs. It then processes the prompt to generate an enhanced version of the original answer, guided exclusively by actionable feedback (\textit{IA}) extracted from user comments and, when necessary, by contextual information from the question.

After processing, the \textit{backend} returns the refined answer to the content script in a standardized JSON format via the background script. The content script then formats the enhanced answer using defined CSS rules, preserving SO’s native prose style and fenced code blocks with optional syntax highlighting, and renders it directly on the page.

\textbf{Performance Comparison of \textit{AUTOCOMBAT} with the Baseline.}
We compared the performance of \textit{AUTOCOMBAT} against the only available baseline, SOUP, recently introduced by Mai et al.~\cite{Mai2024TowardsBA} to enhance SO answers based on comments. As discussed in Section~\ref{sec:introduction}, SOUP is designed to enhance single code snippet based on individual comments and is thus inapplicable to many real-world answers. 
To ensure a fair comparison, we restricted the evaluation to answers in the \textit{ReSOlve} dataset that contain a single code snippet.
Moreover, for answers with multiple comments, we combined all the comments into a single consolidated input before providing them to SOUP. We then applied both \textit{AUTOCOMBAT} and SOUP to enhance these answers in response to the concerns raised in the comments. 
Finally, we evaluated the improvements by comparing the enhanced snippets against human-revised reference answers. We measured the performance of each approach using the same set of syntactic and semantic metrics described in Section~\ref{subsec:human-comparable-ans-improvement}.

\subsection{Developers' Perceptions to Assess the Effectiveness of \textit{AUTOCOMBAT} (RQ3)}

To assess software practitioners' perceptions of \textit{AUTOCOMBAT} 's effectiveness, we conducted a survey following the personal opinion survey guidelines by Kitchenham and Pfleeger \cite{Kitchenham2008PersonalOS}. We also adhered to best practices and ethical standards from prior works \cite{groves2011survey, Singer2002EthicalII}, ensuring informed consent, confidentiality, and a clear explanation of the survey's purpose.
Initially, we conducted a pilot study with five practitioners to evaluate the time required and the clarity of the tasks. Based on their feedback, we made minor adjustments. However, these responses were excluded from the final analysis. Participants were informed that the survey would take approximately 20–25 minutes. Notably, the survey underwent several expert revisions and received approval from University of Saskatchewan's Behavioral Research Ethics Board. The final survey included the following sections.

\noindent\textbf{Consent and Prerequisite.} 
This section asked participants to provide consent, agree to data processing, and confirm their eligibility as SO users with at least one year of experience in Python, Java, C\#, or JavaScript.

\noindent\textbf{Participant Information.} This section collected demographic details, including participants’ country, organization, primary role, development experience, and SO profile age.

\noindent\textbf{\textit{AUTOCOMBAT} Installation and Usage.} 
Participants watched two instructional videos: one on installing \textit{AUTOCOMBAT} and another on using it to enhance SO answers. They then rated the installation ease on a 5-point Likert scale (very easy to very difficult) and assessed the tool’s potential to reduce editing effort and time.


\noindent\noindent\textbf{Evaluation of \textit{AUTOCOMBAT}.} 
Participants evaluated the enhanced answers generated by \textit{AUTOCOMBAT}. Each evaluation included the question, its original answer, and the tool-generated enhancement. Participants reviewed two randomly selected examples in their preferred programming language and rated the enhanced answers on three dimensions, \textit{coverage}, \textit{clarity}, and \textit{relevance}, using a 5-point Likert scale. \textit{Coverage} measured factual accuracy and how well the improvements addressed comment concerns, \textit{clarity} assessed readability and preservation of the original structure, and \textit{relevance} evaluated alignment with the question and initial answer. Finally, participants shared overall impressions of \textit{AUTOCOMBAT}, including whether they would use or recommend it, and provided open-ended feedback for further improvements.


\begin{table}[htb]
\centering
\caption{Experience, profession, and Stack Overflow profile age of the 58 survey participants.}
\resizebox{0.88\textwidth}{!}{
\begin{tabular}{@{}ccc|ccc|ccc@{}}
\toprule
\multicolumn{3}{c|}{\textbf{Development Experience}} & \multicolumn{3}{c|}{\textbf{Profession}} & \multicolumn{3}{c}{\textbf{SO Profile Age}} \\
\cmidrule(r){1-3} \cmidrule(lr){4-6} \cmidrule(l){7-9}
1--3 Yrs & 4--6 Yrs & >6 Yrs & SW Dev. & Aca. & Student & 1--2 Yrs & 3--5 Yrs & >5 Yrs \\
\midrule
55.17\% & 36.21\% & 8.62\% & 52.63\% & 43.85\% & 3.52\% & 33.33\% & 38.60\% & 28.07\% \\
\bottomrule
\end{tabular}
}
\label{tab:survey_participants}
\end{table}

\noindent\textbf{Participants.} We recruited 58 active SO users as participants who satisfied our constraints as follows.

    $\bullet$ \emph{Snowball Approach:} We used convenience sampling to bootstrap the snowball \cite{stratton2021population}. First, we contacted a few software developers who were known to us, easily reachable, and working in software companies worldwide. We explained our study goals and shared the online survey with them. We then adopted a snowballing method \cite{bi2021accessibility} to disseminate the survey to several of their colleagues with similar experiences.
    
    $\bullet$ \emph{Open Circular:} We circulated the survey to specialized Facebook groups. In particular, we targeted the groups where professional software developers discuss their programming problems. We also used \textit{LinkedIn} to find potential participants because it is one of the largest professional networks.

Through these efforts, we recruited participants from diverse geographic regions across North America, Europe, and Asia. Table \ref{tab:survey_participants} summarizes their demographic and professional backgrounds (details can be found in our online appendix \cite{replicationPackage}).

\begin{table}[h]
\centering
\caption{Performance of LLMs in identifying improvement-related comment concerns across quartiles (Q1–Q4)}
\label{tab:rq1}
\resizebox{0.9\textwidth}{!}{
\begin{tabular}{llrrrrrr}
\toprule
Quartile & Model & Accuracy & Precision & Recall & F1 & Specificity & MCC \\
\midrule
\multirow{4}{*}{Q1}
 & DeepSeek & 0.7869 & \textbf{0.7344} & 0.8393 & 0.7833 & \textbf{0.7424} & 0.5805 \\
 & GPT      & 0.7869 & 0.6923 & \textbf{0.9643} & \textbf{0.8060} & 0.6364 & \textbf{0.6233} \\
 & Gemini   & \textbf{0.7951} & 0.7121 & 0.8868 & 0.7899 & 0.7246 & 0.6082 \\
 & LLaMA    & 0.5492 & 0.4831 & 0.8269 & 0.6099 & 0.3429 & 0.1890 \\
\midrule
\multirow{4}{*}{Q2}
 & DeepSeek & 0.7816 & 0.7909 & 0.7982 & 0.7945 & \textbf{0.7629} & 0.5614 \\
 & GPT      & 0.8010 & 0.7568 & \textbf{0.9573} & \textbf{0.8453} & 0.5955 & 0.6088 \\
 & Gemini   & \textbf{0.8107} & \textbf{0.7967} & 0.8750 & 0.8340 & 0.7340 & \textbf{0.6185} \\
 & LLaMA    & 0.6845 & 0.6556 & 0.8839 & 0.7529 & 0.4468 & 0.3724 \\
\midrule
\multirow{4}{*}{Q3}
 & DeepSeek & \textbf{0.8304} & \textbf{0.8115} & 0.8684 & \textbf{0.8390} & \textbf{0.7909} & \textbf{0.6619} \\
 & GPT      & 0.7634 & 0.7207 & \textbf{0.9773} & 0.8296 & 0.4565 & 0.5326 \\
 & Gemini   & 0.8036 & 0.8106 & 0.8492 & 0.8295 & 0.7449 & 0.5991 \\
 & LLaMA    & 0.6339 & 0.6273 & 0.8211 & 0.7113 & 0.4059 & 0.2513 \\
\midrule
\multirow{4}{*}{Q4}
 & DeepSeek & \textbf{0.7731} & \textbf{0.6875} & 0.9167 & \textbf{0.7857} & \textbf{0.6538} & \textbf{0.5810} \\
 & GPT      & 0.6933 & 0.6436 & \textbf{0.9924} & 0.7808 & 0.3271 & 0.4435 \\
 & Gemini   & 0.7185 & 0.6323 & 0.9074 & 0.7452 & 0.5615 & 0.4899 \\
 & LLaMA    & 0.6471 & 0.5941 & 0.8707 & 0.7063 & 0.4344 & 0.3376 \\
\bottomrule
\end{tabular}}
\end{table}

\section{Study Findings}
\label{sec:results}

\subsection{Answering RQ1: Identifying Improvement-Related Concerns in Comments with LLMs}

Table \ref{tab:rq1} reports LLM performance in identifying improvement-related concerns across quartiles (Q1–Q4).

\noindent\textbf{Model-Level Observations.} DeepSeek and GPT are the strongest performers overall. GPT achieves the highest recall across all quartiles, reaching 0.96 in Q1 and climbing as high as 0.99 in Q4, meaning it rarely misses improvement concerns highlighted in comments. This strength makes GPT suitable for comprehensive concern detection. However, its specificity is considerably lower, dropping to 0.33 in Q4, which shows that it frequently flags comments as concerns even when they are not, potentially reducing practical reliability. DeepSeek delivers a more balanced performance: its recall remains high (around 0.84–0.92 across quartiles) while its specificity (up to 0.79 in Q3) and MCC (as high as 0.66 in Q3) are consistently stronger than GPT, making it more dependable. Gemini achieves competitive accuracy (around 0.79–0.81) and F1-scores (up to 0.83 in Q2–Q3), though slightly below DeepSeek and GPT, and is less reliable in distinguishing true concerns from irrelevant ones. LLaMA consistently underperforms across all metrics, with low specificity (as low as 0.34 in Q1) and MCC (dropping to 0.19 in Q1), suggesting that it struggles to separate actionable concerns from background or conversational comments.


\begin{table}[!htbp]
\centering
\caption{Syntactic quality metrics of LLM-enhanced SO answers across quartiles (Q1-Q4).}
\label{tab:llms_eval_syntactic}
\rotatebox{90}{%
\resizebox{0.95\textheight}{!}{
\begin{tabular}{llrrrrrrrrrrrrrrr}
\toprule
Quartile & Model & ROUGE-1 & ROUGE-2 & ROUGE-L & BLEU-1 & BLEU-2 & BLEU-3 & BLEU-4 & METEOR & TER & SacreBLEU & chrF & Jaccard & Dist-1 & Dist-2 & TF-IDF \\
\midrule
\multirow{4}{*}{Q1}
 & DeepSeek & \textbf{0.8423} & \textbf{0.8061} & \textbf{0.8223} & \textbf{0.8234} & \textbf{0.8108} & \textbf{0.8027} & \textbf{0.7965} & \textbf{0.8021} & \textbf{26.4803} & \textbf{82.8238} & \textbf{85.2070} & \textbf{0.7657} & 0.3530 & 0.8005 & \textbf{0.8717} \\
 & GPT      & 0.8065 & 0.7580 & 0.7803 & 0.7533 & 0.7328 & 0.7209 & 0.7121 & 0.7977 & 42.4708 & 73.9253 & 83.4183 & 0.7225 & 0.3426 & 0.8021 & 0.8394 \\
 & Gemini   & 0.8060 & 0.7606 & 0.7828 & 0.8025 & 0.7838 & 0.7723 & 0.7633 & 0.7642 & 36.5956 & 78.0906 & 82.1076 & 0.7210 & 0.3521 & \textbf{0.8030} & 0.8370 \\
 & LLaMA    & 0.7600 & 0.7036 & 0.7315 & 0.7463 & 0.7265 & 0.7139 & 0.7043 & 0.7101 & 38.0472 & 74.8870 & 78.2668 & 0.6584 & \textbf{0.3570} & 0.7994 & 0.7946 \\
\midrule
\multirow{4}{*}{Q2}
 & DeepSeek & \textbf{0.7657} & \textbf{0.7079} & \textbf{0.7379} & 0.6705 & 0.6505 & 0.6377 & 0.6279 & \textbf{0.6827} & \textbf{47.6312} & 65.1665 & 72.3509 & \textbf{0.6525} & 0.2881 & \textbf{0.7511} & \textbf{0.8155} \\
 & GPT      & 0.7119 & 0.6288 & 0.6664 & \textbf{0.7071} & \textbf{0.6746} & \textbf{0.6550} & \textbf{0.6405} & 0.6554 & 54.4688 & \textbf{68.4075} & \textbf{72.5614} & 0.5624 & 0.2791 & 0.7490 & 0.7654 \\
 & Gemini   & 0.7308 & 0.6584 & 0.6914 & 0.6817 & 0.6503 & 0.6310 & 0.6167 & 0.6525 & 52.4996 & 66.1466 & 70.9200 & 0.5945 & 0.2853 & 0.7426 & 0.7830 \\
 & LLaMA    & 0.7152 & 0.6406 & 0.6783 & 0.6236 & 0.5988 & 0.5827 & 0.5704 & 0.6239 & 51.9765 & 62.4086 & 69.0492 & 0.5748 & \textbf{0.2944} & 0.7452 & 0.7731 \\
\midrule
\multirow{4}{*}{Q3}
 & DeepSeek & \textbf{0.7456} & \textbf{0.6824} & \textbf{0.7161} & 0.6591 & \textbf{0.6393} & \textbf{0.6268} & \textbf{0.6172} & \textbf{0.6704} & \textbf{50.4759} & 65.3750 & \textbf{72.3385} & \textbf{0.6223} & \textbf{0.3052} & \textbf{0.7705} & \textbf{0.8177} \\
 & GPT      & 0.6786 & 0.5896 & 0.6301 & \textbf{0.6695} & 0.6330 & 0.6122 & 0.5972 & 0.6324 & 70.4968 & \textbf{66.4256} & 70.6501 & 0.5159 & 0.2889 & 0.7629 & 0.7625 \\
 & Gemini   & 0.7152 & 0.6446 & 0.6805 & 0.6650 & 0.6382 & 0.6212 & 0.6084 & 0.6377 & 56.9894 & 65.6391 & 70.2565 & 0.5759 & 0.2992 & 0.7652 & 0.7883 \\
 & LLaMA    & 0.6854 & 0.6093 & 0.6522 & 0.6039 & 0.5823 & 0.5681 & 0.5570 & 0.5967 & 55.4440 & 60.0376 & 68.2936 & 0.5336 & 0.3060 & 0.7677 & 0.7752 \\
\midrule
\multirow{4}{*}{Q4}
 & DeepSeek & \textbf{0.7188} & \textbf{0.6410} & \textbf{0.6786} & 0.5549 & 0.5309 & 0.5156 & 0.5040 & \textbf{0.6318} & \textbf{61.5339} & 54.3388 & \textbf{64.1222} & \textbf{0.5715} & \textbf{0.2943} & \textbf{0.7721} & \textbf{0.8016} \\
 & GPT      & 0.6373 & 0.5330 & 0.5766 & \textbf{0.6123} & \textbf{0.5693} & \textbf{0.5441} & \textbf{0.5259} & 0.5820 & 80.3163 & \textbf{57.8818} & 63.5569 & 0.4605 & 0.2716 & 0.7551 & 0.7422 \\
 & Gemini   & 0.6708 & 0.5738 & 0.6187 & 0.5806 & 0.5426 & 0.5187 & 0.5008 & 0.5860 & 72.1832 & 55.3928 & 62.3221 & 0.4967 & 0.2825 & 0.7568 & 0.7653 \\
 & LLaMA    & 0.6762 & 0.5989 & 0.6378 & 0.5239 & 0.5002 & 0.4847 & 0.4729 & 0.5912 & 65.3996 & 51.4111 & 61.4159 & 0.5214 & 0.2942 & 0.7664 & 0.7775 \\
\bottomrule
\end{tabular}}}
\end{table}

\begin{table}[h]
\centering
\caption{Semantic quality metrics of LLM-enhanced SO answers across quartiles (Q1–Q4).}
\label{tab:llms_eval_semantic}
\resizebox{0.9\textwidth}{!}{
\begin{tabular}{llrrrrr}
\toprule
Quartile & Model & BLEURT & COMET & BARTScore & BERTScore\_F1 & MoverScore \\
\midrule
\multirow{4}{*}{Q1}
 & DeepSeek & \textbf{0.7767} & \textbf{0.8231} & -1.6435 & \textbf{0.7707} & \textbf{0.9034} \\
 & GPT      & 0.7503 & 0.8177 & \textbf{-1.5839} & 0.7362 & 0.8866 \\
 & Gemini   & 0.7567 & 0.8146 & -1.7479 & 0.7342 & 0.8785 \\
 & LLaMA    & 0.7059 & 0.7942 & -1.9949 & 0.6353 & 0.8505 \\
\midrule
\multirow{4}{*}{Q2}
 & DeepSeek & \textbf{0.7159} & \textbf{0.7928} & \textbf{-1.8108} & \textbf{0.5973} & \textbf{0.8633} \\
 & GPT      & 0.6620 & 0.7763 & -1.8656 & 0.5211 & 0.8257 \\
 & Gemini   & 0.6819 & 0.7808 & -1.9404 & 0.5393 & 0.8302 \\
 & LLaMA    & 0.6680 & 0.7742 & -2.0127 & 0.5132 & 0.8272 \\
\midrule
\multirow{4}{*}{Q3}
 & DeepSeek & \textbf{0.6995} & \textbf{0.7837} & \textbf{-1.8581} & \textbf{0.4938} & \textbf{0.8613} \\
 & GPT      & 0.6320 & 0.7713 & -1.9379 & 0.4109 & 0.8181 \\
 & Gemini   & 0.6669 & 0.7716 & -1.9449 & 0.4347 & 0.8311 \\
 & LLaMA    & 0.6507 & 0.7643 & -2.1120 & 0.4014 & 0.8262 \\
\midrule
\multirow{4}{*}{Q4}
 & DeepSeek & \textbf{0.6686} & \textbf{0.7770} & \textbf{-1.9789} & \textbf{0.4732} & \textbf{0.8443} \\
 & GPT      & 0.5967 & 0.7609 & -2.0794 & 0.3714 & 0.7944 \\
 & Gemini   & 0.6181 & 0.7603 & -2.1105 & 0.3598 & 0.7986 \\
 & LLaMA    & 0.6416 & 0.7611 & -2.0753 & 0.4065 & 0.8221 \\
\bottomrule
\end{tabular}}
\end{table}

\noindent\textbf{Quartile-Level Observations.} Performance also varies across quartiles, which capture the distribution of comment counts. In Q1 (answers with a single comment), GPT and DeepSeek already perform strongly, suggesting that even limited feedback can signal improvement opportunities. However, LLaMA struggles significantly here, showing that sparse feedback makes it harder for weaker models to interpret intent. In Q2 and Q3 (answers with 2–5 comments), all models improve, with DeepSeek achieving its best balance of precision and recall. These quartiles appear to provide enough feedback for models to reliably identify concerns without overwhelming noise. Interestingly, in Q4 (answers with 6–13 comments), recall remains high for GPT and DeepSeek, but specificity drops across all models, indicating that as answers accumulate more comments, the mix of relevant and off-topic remarks increases, making it harder for models to filter true concerns. 


\begin{findingbox}
\textbf{RQ\textsubscript{1} Summary}: LLMs are able to detect improvement-related concerns in comments, but their effectiveness differs across models and feedback volumes. GPT captures almost all concerns but tends to over-flag, while DeepSeek offers balanced and reliable performance. Gemini performs moderately well, whereas LLaMA struggles consistently. Models perform optimally with a moderate number of comments, as very high comment counts introduce noise that reduces accuracy.
\end{findingbox}

\subsection{Answering RQ2: Assessing LLM's Human-Comparable Answer Improvements}
\label{rq2}

\subsubsection{Effectiveness Evaluation of \textit{AUTOCOMBAT}} 

Table \ref{tab:llms_eval_syntactic} and Table \ref{tab:llms_eval_semantic} present syntactic and semantic evaluation measures of the \textit{AUTOCOMBAT}, which leverages the DeepSeek model.

\noindent\textbf{Model-Level Observations.} DeepSeek consistently delivers the strongest revisions across both syntactic and semantic metrics. For instance, in Q1 it achieves the highest ROUGE-L (0.82) and BLEURT (0.78), while also leading COMET (0.82), showing strong alignment with human revisions. GPT is a close competitor, reaching comparable semantic scores (e.g., BLEURT 0.75, COMET 0.81 in Q1) but with lower syntactic precision (ROUGE-L 0.78). Gemini performs moderately, with balanced results but consistently below DeepSeek (e.g., BLEURT around 0.76 in Q1, dropping to 0.62 in Q4). LLaMA lags behind, with substantially lower semantic alignment (e.g., BLEURT 0.71 in Q1 and 0.64 in Q4), reflecting its weaker ability to integrate feedback into meaningful improvements.


\noindent\textbf{Quartile-Level Observations.} Performance patterns across quartiles reveal how the number of comments influences refinement quality. In Q1 (single-comment answers), models produce the strongest improvements, with DeepSeek leading both syntactically and semantically. Q2 and Q3 (2–5 comments) yield moderately strong results: for example, DeepSeek achieves COMET values of 0.78–0.79 and maintains ROUGE-L around 0.74, indicating that it can still leverage moderate feedback effectively. In contrast, Q4 (6–13 comments) shows a drop across models, with BLEURT scores falling for GPT (0.59) and Gemini (0.62), and syntactic measures like ROUGE-L declining below 0.64 for most models. This pattern suggests that higher comment volumes introduce conflicting or irrelevant feedback, making it harder for models to identify which suggestions are actionable.

\noindent\textbf{Cross-RQ Insights.} The findings for RQ2 complement those from RQ1. In RQ1, GPT achieved very high recall (up to 0.99) but often over-flagged irrelevant concerns, while DeepSeek offered a stronger balance with higher specificity and MCC. This trend carries into RQ2: GPT tends to over-edit, introducing changes that boost recall-like metrics but sometimes reduce syntactic precision, whereas DeepSeek translates its balanced detection into balanced, human-comparable revisions. Thus, a model’s ability to reliably identify concerns (RQ1) directly influences the quality and faithfulness of its revisions (RQ2).

\begin{figure*}[h] 
    \centering
    \footnotesize 

    \begin{subfigure}[b]{0.32\linewidth} 
        \centering
        \togglefalse{showpct}
        \resizebox{\linewidth}{!}{
            \begin{tikzpicture} 
                \pie[explode={0.1, 0.1, 0.1, 0.2}, radius=1.0, text=pin, color={black!50, gray!45, gray!20, white}]{
                25.9/Very Easy (25.9\%),
                44.8/Easy (44.8\%),
                24.1/Moderate (24.1\%),
                5.1/Difficult (5.1\%)
                }
            \end{tikzpicture}
        }
        \caption{Ease of installation}
        \label{fig:installation}
    \end{subfigure}%
    \hfill 
    \begin{subfigure}[b]{0.32\linewidth} 
        \centering
        \togglefalse{showpct}
        \resizebox{\linewidth}{!}{
            \begin{tikzpicture} 
                \pie[explode={0.1, 0.1, 0.2}, radius=1.0, text=pin, color={black!50, gray!20, white}]{
                48.3/Significantly (48.3\%),
                39.7/Moderately (39.7\%),
                11.0/Slightly (11.0\%)
                }
            \end{tikzpicture}
        }
        \caption{Effort and time reduction}
        \label{fig:effort}
    \end{subfigure}%
    \hfill 
    \begin{subfigure}[b]{0.42\linewidth}
        \centering
        \resizebox{\linewidth}{!}{
            \begin{tikzpicture} 
                \pie[explode={0.1, 0.1, 0.1, 0.2, 0.2}, radius=1.2, text=pin, color={black!50, gray!45, gray!25, gray!10, white}]{
                37.9/\normalsize Very Likely (37.9\%),
                46.6/\normalsize Somewhat Likely (46.6\%),
                12.1/\normalsize Neutral (12.1\%),
                2.4/\normalsize Somewhat Unlikely (2.4\%)
                }
            \end{tikzpicture}
        }
        \caption{Recommendation likelihood}
        \label{fig:recommendation}
    \end{subfigure}

    \caption{Summary of \textit{AUTOCOMBAT} user feedback: (a) Ease of installation on browser, (b) time and effort reduction for enhancing SO answers, and (c) likelihood of using and recommending \textit{AUTOCOMBAT}.}
    \label{fig:overall_feedback}
\end{figure*}
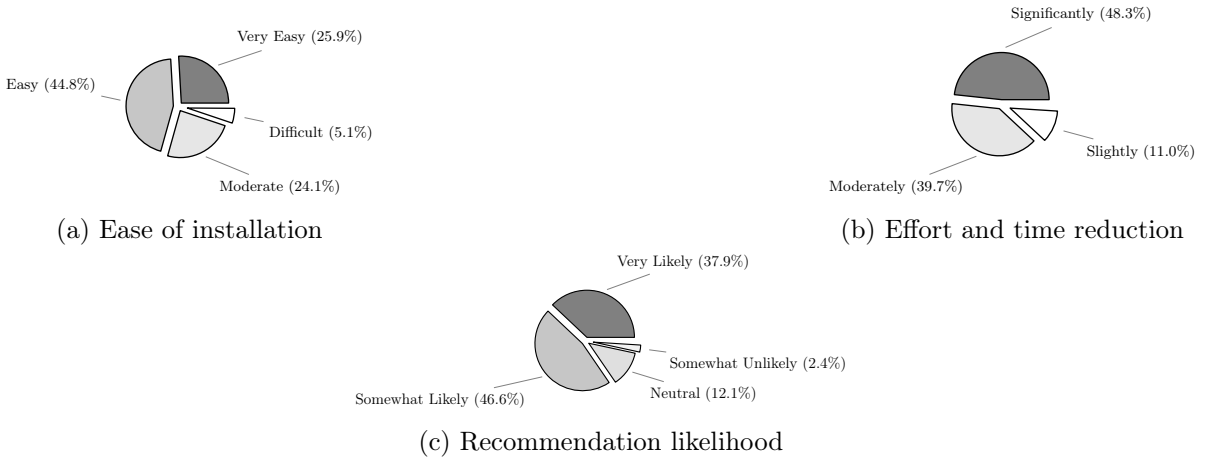

\subsubsection{Analysis of Intent Preservation}
\label{sec:intent_preservation}

Table~\ref{tab:intent_preservation_distribution} presents the percentage distribution of intent preservation outcomes across quartiles (Q1--Q4), where higher quartiles correspond to answers associated with larger comment threads. Overall, the results indicate that most LLM-enhanced answers preserve the original intent, although the frequency and severity of deviations vary across models and levels of interaction.

DeepSeek demonstrates the strongest intent preservation performance across all quartiles. In Q1, where answers are associated with a single comment, 85.25\% of DeepSeek’s refinements fully preserve intent, with only 2.46\% resulting in complete intent violations. This trend becomes more pronounced in higher quartiles. In Q4, DeepSeek achieves the highest preservation rate at 91.60\% and the lowest proportion of complete intent loss at 0.42\%. These results suggest that DeepSeek effectively incorporates user feedback while maintaining alignment with the original technical goal, even as the complexity of comment-driven refinement increases.

\begin{table}[!h]
\centering
\caption{Percentage distribution of intent preservation types across quartiles (Q1--Q4).}
\label{tab:intent_preservation_distribution}
\resizebox{0.7\textwidth}{!}{
\begin{tabular}{llrrr}
\toprule
Quartile & Model & YES (\%) & PARTIALLY YES (\%) & NO (\%) \\
\midrule
\multirow{4}{*}{Q1}
 & DeepSeek & \textbf{85.25} & 12.30 & 2.46 \\
 & GPT      & 84.43 & 13.11 & 2.46 \\
 & Gemini   & 82.79 & 13.11 & \textbf{4.10} \\
 & LLaMA    & 81.97 & \textbf{16.39} & 1.64 \\
\midrule
\multirow{4}{*}{Q2}
 & DeepSeek & \textbf{84.95} & 14.56 & 0.49 \\
 & GPT      & 82.04 & 15.53 & \textbf{2.43} \\
 & Gemini   & 81.07 & \textbf{16.99} & 1.94 \\
 & LLaMA    & 82.52 & 16.02 & 1.46 \\
\midrule
\multirow{4}{*}{Q3}
 & DeepSeek & \textbf{89.73} & 8.93 & 1.34 \\
 & GPT      & 83.48 & 15.62 & 0.89 \\
 & Gemini   & 84.82 & 13.39 & 1.79 \\
 & LLaMA    & 77.68 & \textbf{19.64} & \textbf{2.68} \\
\midrule
\multirow{4}{*}{Q4}
 & DeepSeek & \textbf{91.60} & 7.98 & 0.42 \\
 & GPT      & 85.29 & 14.29 & 0.42 \\
 & Gemini   & 89.50 & 8.82 & \textbf{1.68} \\
 & LLaMA    & 83.61 & \textbf{15.55} & 0.84 \\
\bottomrule
\end{tabular}}
\end{table}

GPT and Gemini exhibit similar but slightly weaker patterns. Both models preserve intent in more than 80\% of cases across all quartiles, but they show consistently higher rates of partial intent deviations. For GPT, partially preserved intent reaches 15.62\% in Q3, indicating a tendency to reinterpret or extend feedback in ways that subtly alter scope. Gemini follows a comparable trend, with partial intent preservation peaking at 16.99\% in Q2 and complete intent violations reaching 4.10\% in Q1. These findings suggest that while GPT and Gemini usually retain the core goal of the original answer, they are more prone to introducing additional assumptions or broader changes when responding to feedback.

LLaMA exhibits the weakest intent preservation among the evaluated models. Although its rate of complete intent loss remains relatively low, it consistently shows the highest proportion of partial intent deviations, reaching 19.64\% in Q3 and remaining above 15\% in all quartiles. This pattern indicates difficulty in balancing multiple comment-driven concerns while maintaining strict alignment with the original question context.

Across quartiles, intent preservation generally improves as the number of comments increases, particularly for DeepSeek and Gemini. This suggests that richer feedback often provides clearer guidance for refinement, reducing the likelihood of complete misalignment. However, the persistence of partial intent deviations even in Q4 highlights the inherent challenge of integrating multiple user concerns without altering the original technical goal. 


To provide concrete insight into how intent preservation was assessed, we next present illustrative examples drawn from the manual evaluation.

\textbf{Example of Complete Intent Preservation.} The first example demonstrates a case of complete intent preservation. The original question asks how to determine whether an array of integers contains a pair of numbers whose product is odd, which relies on the property that the product of two integers is odd if and only if both integers are odd \cite{StackOverflowOddProduct}. The initial answer, however, incorrectly focused on detecting even numbers, as shown below.

\begin{lstlisting}
public static boolean hasAtLeastTwoEvens(int[] args) {
    int[] target = args; // make defensive copy
    int evensFound;
    int numberEvensSought = 2;

    for (int i = 0; i < target.length; i++) {
        if (target[i] % 2 == 0) {
            if (evensFound == numberEvensSought) {
                return true;
            }
            evensFound++;
        }
    }

    return false;
}    
\end{lstlisting}

This implementation fails to address the core intent of the question. The associated comment thread explicitly highlights this mismatch by noting that the solution should reason about odd numbers rather than even ones. In response, the human-revised reference answer corrects the parity condition and updates the function and variable names to reflect the intended goal, while preserving the original brute force structure of the algorithm.

\begin{lstlisting}
public static boolean hasAtLeastTwoOdds(int[] args) {
    int[] target = args; // make defensive copy
    int oddsFound;
    int numberOddsSought = 2;

    for (int i = 0; i < target.length; i++) {
        if (target[i] % 2 != 0) {
            if (oddsFound == numberOddsSought) {
                return true;
            }
            oddsFound++;
        }
    }

    return false;
}
\end{lstlisting}

The DeepSeek-generated enhanced answer preserves the original technical intent completely by producing a revision that is semantically and structurally equivalent to the human-revised answer. It applies the correct parity check for odd integers, maintains the same stopping condition, and follows the same algorithmic logic without introducing additional assumptions or altering the scope of the solution.




Because the DeepSeek-generated revision addresses the same problem under the same assumptions and integrates the comment feedback without modifying the original goal, this case was labeled as \emph{YES} in our manual intent preservation analysis. 

\textbf{Example of Partial Intent Preservation.}
The second example illustrates partial intent preservation in the context of fixing a SQLAlchemy query error in a FastAPI application. The original question reports an \texttt{AttributeError} raised when attempting to access a value from the request body using attribute notation on a Python dictionary in a database query \cite{StackOverflowSqlalchemy}. 

The initial answer correctly identifies the cause of the runtime error and explains that dictionary values in Python must be accessed using square bracket notation rather than dot notation. The answer updates the code accordingly, replacing attribute access with key-based access, as shown below.

\begin{lstlisting}
@app.get("/datasets/")
def get_project_registry(body: Dict[str, Any], db: Session = Depends(get_db)):
    stmt = select(custom_datasets).where(
        custom_datasets.c.project_id in body['project_ids']
    )
    res = db.execute().all()
    return res
\end{lstlisting}

However, a follow-up comment points out that filtering is still not working. The query returns no results even though the table contains a row with the requested \texttt{project\_id}. This indicates that an additional issue remains unresolved.

The human-revised reference answer addresses this concern by ensuring that the constructed statement is executed and by explaining that the Python \texttt{in} operator cannot be used directly inside a SQLAlchemy \texttt{where} clause. Instead, SQLAlchemy requires the use of the \texttt{.in\_()} method to construct a proper SQL \texttt{IN} expression. The revised answer incorporates these corrections, ensuring that the query executes properly and performs correct filtering.

\begin{lstlisting}
@app.get("/datasets/")
def get_project_registry(body: Dict[str, Any], db: Session = Depends(get_db)):
    stmt = select(custom_datasets).where(
        custom_datasets.c.project_id.in_(body['project_ids'])
    )
    res = db.execute(stmt).all()
    return res
\end{lstlisting}

In contrast, the Gemini-generated enhanced answer ensures that the constructed statement is passed to \texttt{db.execute}, thereby resolving the execution issue. However, it does not address the incorrect use of the Python \texttt{in} operator within the SQLAlchemy \texttt{where} clause. As a result, the filtering condition is not translated into a proper SQL \texttt{IN} expression, and the query still fails to retrieve the expected records.




\begin{lstlisting}
@app.get("/datasets/")
def get_project_registry(body: Dict[str, Any], db: Session = Depends(get_db)):
    stmt = select(custom_datasets).where(
        custom_datasets.c.project_id in body['project_ids']
    )
    res = db.execute(stmt).all()
    return res
\end{lstlisting}

Although the Gemini-generated revision preserves the general intent of correcting the reported error, it does not fully incorporate the improvement-related feedback regarding query filtering. The overall technical goal remains identifiable, but an essential constraint for correct filtering behavior is not met. Therefore, this example is labeled as \emph{PARTIALLY YES} in our manual intent preservation analysis.

\subsubsection{Baseline Comparison with \textit{AUTOCOMBAT}} 
We compared our approach with SOUP \cite{Mai2024TowardsBA}, the only existing baseline for comment-driven answer refinement. 
%
Tables~\ref{tab:syntactic} and~\ref{tab:semantic} summarize the results for syntactic and semantic similarity, respectively.
\textit{AUTOCOMBAT} consistently outperformed SOUP across all evaluation metrics. For syntactic similarity, it achieved substantially higher scores (\textit{ROUGE-L = 0.83} vs.\ 0.53, \textit{BLEU-4 = 0.78} vs.\ 0.25, \textit{METEOR = 0.79} vs.\ 0.48) and a lower \textit{TER} value (24.97 vs.\ 66.68), indicating that its refinements required far fewer edits to align with human references. Semantic evaluations showed a similar pattern, with \textit{AUTOCOMBAT} reaching \textit{BLEURT = 0.73}, \textit{COMET = 0.85}, \textit{BERTScore\_F1 = 0.81}, and \textit{MoverScore = 0.91}, all notably higher than SOUP.

\begin{table*}[!h]
\centering
\caption{Comparative analysis of \textit{AUTOCOMBAT} and SOUP on syntactic metrics.}
\label{tab:syntactic}
\resizebox{\textwidth}{!}{
\begin{tabular}{lrrrrrrrrrrrrrrr}
\toprule
Method & ROUGE-1 & ROUGE-2 & ROUGE-L & BLEU-1 & BLEU-2 & BLEU-3 & BLEU-4 & METEOR & TER & SacreBLEU & chrF & Jaccard & Dist-1 & Dist-2 & TF-IDF \\
\midrule
SOUP & 0.5474 & 0.4065 & 0.5296 & 0.3259 & 0.2968 & 0.2733 & 0.2524 & 0.4840 & 66.6756 & 16.4341 & 36.0951 & 0.6208 & 0.4000 & 0.6770 & 0.6770 \\
\cellcolor[HTML]{CCCCCC}\texttt{\textbf{AUTOCOMBAT}} & 
\cellcolor[HTML]{CCCCCC}0.8416 & 
\cellcolor[HTML]{CCCCCC}0.7864 & 
\cellcolor[HTML]{CCCCCC}0.8256 & 
\cellcolor[HTML]{CCCCCC}0.8133 & 
\cellcolor[HTML]{CCCCCC}0.7976 & 
\cellcolor[HTML]{CCCCCC}0.7856 & 
\cellcolor[HTML]{CCCCCC}0.7756 & 
\cellcolor[HTML]{CCCCCC}0.7927 & 
\cellcolor[HTML]{CCCCCC}24.8675 & 
\cellcolor[HTML]{CCCCCC}77.2102 & 
\cellcolor[HTML]{CCCCCC}81.4651 & 
\cellcolor[HTML]{CCCCCC}0.8543 & 
\cellcolor[HTML]{CCCCCC}0.4639 & 
\cellcolor[HTML]{CCCCCC}0.7903 & 
\cellcolor[HTML]{CCCCCC}0.8886 \\
\bottomrule
\end{tabular}} 
\end{table*}


\begin{table}[!h]
\centering
\caption{Comparative analysis of \textit{AUTOCOMBAT} and SOUP on semantic metrics.}
\label{tab:semantic}
\resizebox{0.75\textwidth}{!}{
\begin{tabular}{lrrrrr}
\toprule
Method & BLEURT & COMET & BARTScore & BERTScore\_F1 & MoverScore \\
\midrule
SOUP & 0.5531 & 0.6645 & -2.4000 & 0.3385 & 0.7273 \\
\cellcolor[HTML]{CCCCCC}\texttt{\textbf{AUTOCOMBAT}} & 
\cellcolor[HTML]{CCCCCC}0.7299 & 
\cellcolor[HTML]{CCCCCC}0.8481 & 
\cellcolor[HTML]{CCCCCC}-1.0260 & 
\cellcolor[HTML]{CCCCCC}0.8139 & 
\cellcolor[HTML]{CCCCCC}0.9126\\
\bottomrule
\end{tabular}}
\end{table}

We further conducted a \emph{Mann-Whitney-Wilcoxon} \cite{mcknight2010mann}, a non-parametric test to examine whether these differences were statistically significant, and the results confirmed significance (p-value $\approx$ 0.0 $<$ 0.05) across all syntactic and semantic metrics, indicating that the improvements achieved by \textit{AUTOCOMBAT} are not only consistent but also statistically significant. This advantage stems from its ability to process multiple concerns simultaneously, producing coherent and balanced revisions that capture diverse perspectives. Moreover, by selectively incorporating question context, applied in about 31\% of enhancement cases, \textit{AUTOCOMBAT} effectively resolves ambiguity and preserves the original intent of the answer, resulting in refinements that are both semantically accurate and contextually precise.



\begin{findingbox}
\textbf{RQ\textsubscript{2} Summary}: \textit{AUTOCOMBAT} produces revisions that closely resemble human edits by effectively integrating multiple comment concerns and selectively incorporating question context. It performs strongest on answers with a small to moderate number of comments and consistently outperforms SOUP and other LLMs, delivering clearer and more contextually aligned refinements. Performance declines for answers with very high comment volumes, where conflicting or irrelevant feedback makes refinement more challenging. While GPT and Gemini achieve competitive results, they are less stable, and LLaMA consistently lags behind across all evaluation metrics. Additionally, manual intent analysis confirms that \textit{AUTOCOMBAT} preserves the original technical goal more reliably than competing models, with fewer partial or complete intent deviations across quartiles.
\end{findingbox}

 

\subsection{Answering RQ3: User Perception of \textit{AUTOCOMBAT}-Enhanced Answers in Practical Settings}
\label{rq3}

\noindent\textbf{\textit{AUTOCOMBAT} Installation and Usage.} 
Among 58 participants, the majority found the installation process straightforward. As shown in Fig. \ref{fig:installation}, 25.9\% marked it as \textit{“Very Easy”} and 44.8\% as \textit{“Easy.”} A smaller portion, 24.1\%, considered it \textit{“Moderate,”} while only 5.1\% found it \textit{“Difficult.”}
Fig. \ref{fig:effort} shows that \textit{AUTOCOMBAT} substantially reduced participants’ effort and time in editing SO answers, with 48.3\% describing the reduction as \textit{“Significant,”} 39.7\% as \textit{“Moderate,”} and 11.0\% as \textit{“Slight.”} These results suggest that \textit{AUTOCOMBAT} effectively streamlines the answer improvement process, enabling users to enhance answers more efficiently and with less manual effort.

\noindent\textbf{Language-wise Evaluation of \textit{AUTOCOMBAT}.} Answer quality was assessed across \textit{coverage}, \textit{clarity}, and \textit{relevance} (Table \ref{tab:rq4_evalfindings}). Python received the highest ratings (4.3 for \textit{coverage} and \textit{clarity}, 4.5 for \textit{relevance}), with 58.8\% rating \textit{relevance} as \textit{“Excellent”} (5). Java followed closely (4.1 for \textit{coverage} and \textit{clarity}, 4.2 for \textit{relevance}), with 46.7\% giving the highest rating for relevance. JavaScript and C\# had slightly lower ratings but remained effective. JavaScript averaged 3.8 for \textit{coverage} and \textit{clarity} and 4.0 for \textit{relevance}, while C\# scored 3.8, 3.9, and 3.9, respectively.

\begin{table}[!htbp]
\caption{Language-wise evaluation findings for answers enhanced by \textit{AUTOCOMBAT} among survey participants (Here, \textbf{M}: \textit{Mean}, \textbf{Med}: \textit{Median}. Ratings (\textbf{5}, \textbf{4}, \textbf{3}, \textbf{2}, \textbf{1}) represent \textit{Excellent}, \textit{Very Good}, \textit{Good}, \textit{Bad}, and \textit{Very Bad} respectively).}
\label{tab:rq4_evalfindings}
\resizebox{\textwidth}{!}{%
\begin{tabular}{@{}l|cccccc|ccccccccccccccc@{}}
\toprule
\multicolumn{1}{c|}{\multirow{3}{*}{\textbf{\begin{tabular}[c]{@{}c@{}}Programming\\ Language\end{tabular}}}} & \multicolumn{6}{c|}{\textbf{Overall Evaluation Score}}                                                                                          & \multicolumn{15}{c}{\textbf{Individual Ratings (\% of Participants)}}                                                                                                                                                                      \\ \cmidrule(l){2-22} 
\multicolumn{1}{c|}{}                                                                                         & \multicolumn{2}{c|}{\textbf{Coverage}}         & \multicolumn{2}{c|}{\textbf{Clarity}}          & \multicolumn{2}{c|}{\textbf{Relevance}} & \multicolumn{5}{c|}{\textbf{Coverage}}                                              & \multicolumn{5}{c|}{\textbf{Clarity}}                                               & \multicolumn{5}{c}{\textbf{Relevance}}                         \\ \cmidrule(l){2-22} 
\multicolumn{1}{c|}{}                                                                                         & \textbf{M} & \multicolumn{1}{c|}{\textbf{Med}} & \textbf{M} & \multicolumn{1}{c|}{\textbf{Med}} & \textbf{M}        & \textbf{Med}        & \textbf{5} & \textbf{4} & \textbf{3} & \textbf{2} & \multicolumn{1}{c|}{\textbf{1}} & \textbf{5} & \textbf{4} & \textbf{3} & \textbf{2} & \multicolumn{1}{c|}{\textbf{1}} & \textbf{5} & \textbf{4} & \textbf{3} & \textbf{2} & \textbf{1} \\ \midrule
\textbf{Python}                                                                                               & 4.3        & \multicolumn{1}{c|}{4.5}          & 4.3        & \multicolumn{1}{c|}{4.0}          & 4.5               & 5.0                 & 50.0       & 29.4       & 17.6       & 2.9        & \multicolumn{1}{c|}{-}          & 45.6       & 38.2       & 14.7       & 1.5        & \multicolumn{1}{c|}{-}          & 58.8       & 29.4       & 11.8       & -          & -          \\
\textbf{Java}                                                                                                 & 4.1        & \multicolumn{1}{c|}{4.0}          & 4.1        & \multicolumn{1}{c|}{4.0}          & 4.2               & 4.0                 & 33.3       & 40.0       & 23.3       & 3.3        & \multicolumn{1}{c|}{-}          & 33.3       & 40.0       & 23.3       & 3.3        & \multicolumn{1}{c|}{-}          & 46.7       & 30.0       & 23.3       & -          & -          \\
\textbf{JavaScript}                                                                                           & 3.8        & \multicolumn{1}{c|}{4.0}          & 3.8        & \multicolumn{1}{c|}{4.0}          & 4.0               & 4.0                 & -          & 75.0       & 25.0       & -          & \multicolumn{1}{c|}{-}          & -          & 75.0       & 25.0       & -          & \multicolumn{1}{c|}{-}          & 25.0       & 50.0       & 25.0       & -          & -          \\
\textbf{C\#}                                                                                                  & 3.8        & \multicolumn{1}{c|}{3.5}          & 3.9        & \multicolumn{1}{c|}{4.0}          & 3.9               & 4.0                 & 28.6       & 21.4       & 50.0       & -          & \multicolumn{1}{c|}{-}          & 28.6       & 35.7       & 35.7       & -          & \multicolumn{1}{c|}{-}          & 21.4       & 50.0       & 28.6       & -          & -          \\ \bottomrule
\end{tabular}%
}
\end{table}

\noindent\textbf{Profession and Experience-wise Evaluation of \textit{AUTOCOMBAT}.} 
We further analyzed the ratings based on participants’ professions and experience levels, as shown in Table \ref{tab:rq4_profexp_eval}. Academicians provided the highest ratings, with a median score of 5.0 across most metrics and 57.7\% rating \textit{relevance} as \textit{“Excellent.”} Software developers also responded positively, with median scores of 4.0 across all metrics, frequently rating enhancements as \textit{“Excellent”} or \textit{“Very Good.”} Students gave slightly lower ratings, with median scores around 4.0 and fewer \textit{“Excellent”} responses.

According to experience, participants with over four years of experience assigned median scores of 5.0 to most metrics, while those with more than six years of experience reported slightly lower clarity ratings (4.5) but continued to value \textit{coverage} and \textit{relevance} highly, with over 58\% rating them as \textit{“Excellent.”}
These results indicate that \textit{AUTOCOMBAT} is perceived as highly effective across diverse user groups, regardless of profession or experience level.

\begin{table}[h]
\caption{Profession and experience-wise evaluation findings for answers enhanced by \textit{AUTOCOMBAT} among survey participants. (Here, \textbf{M}: \textit{Mean}, \textbf{Med}: \textit{Median}. Ratings (\textbf{5}, \textbf{4}, \textbf{3}, \textbf{2}, \textbf{1}) represent \textit{Excellent}, \textit{Very Good}, \textit{Good}, \textit{Bad}, and \textit{Very Bad} respectively).}
\label{tab:rq4_profexp_eval}

\centering
\resizebox{\textwidth}{!}{%
\begin{tabular}{@{}ll|cccccc|ccccccccccccccc@{}}
\toprule
\multicolumn{2}{c|}{\multirow{3}{*}{\textbf{Criteria}}}                                   & \multicolumn{6}{c|}{\textbf{Overall Evaluation Score}}                                                                                    & \multicolumn{15}{c}{\textbf{Individual Ratings (\% of Participants)}}                                                                                                                                                                      \\ \cmidrule(l){3-23} 
\multicolumn{2}{c|}{}                                                                     & \multicolumn{2}{c|}{\textbf{Coverage}}         & \multicolumn{2}{c|}{\textbf{Clarity}}          & \multicolumn{2}{c|}{\textbf{Relevance}} & \multicolumn{5}{c|}{\textbf{Coverage}}                                              & \multicolumn{5}{c|}{\textbf{Clarity}}                                               & \multicolumn{5}{c}{\textbf{Relevance}}                         \\ \cmidrule(l){3-23} 
\multicolumn{2}{c|}{}                                                                     & \textbf{M} & \multicolumn{1}{c|}{\textbf{Med}} & \textbf{M} & \multicolumn{1}{c|}{\textbf{Med}} & \textbf{M}        & \textbf{Med}        & \textbf{5} & \textbf{4} & \textbf{3} & \textbf{2} & \multicolumn{1}{c|}{\textbf{1}} & \textbf{5} & \textbf{4} & \textbf{3} & \textbf{2} & \multicolumn{1}{c|}{\textbf{1}} & \textbf{5} & \textbf{4} & \textbf{3} & \textbf{2} & \textbf{1} \\ \midrule
\multicolumn{1}{l|}{\multirow{3}{*}{\textbf{Profession}}} & \textbf{Academician}          & 4.2        & \multicolumn{1}{c|}{5.0}          & 4.3        & \multicolumn{1}{c|}{4.5}          & 4.4               & 5.0                 & 51.9       & 21.2       & 23.1       & 3.8        & \multicolumn{1}{c|}{-}          & 50         & 30.8       & 19.2       & -          & \multicolumn{1}{c|}{-}          & 57.7       & 26.9       & 15.4       & -          & -          \\
\multicolumn{1}{l|}{}                                     & \textbf{SW Developer}         & 4.1        & \multicolumn{1}{c|}{4.0}          & 4.1        & \multicolumn{1}{c|}{4.0}          & 4.2               & 4.0                 & 38.5       & 42.3       & 1.5        & 5.8        & \multicolumn{1}{c|}{-}          & 36.5       & 38.5       & 19.2       & 5.8        & \multicolumn{1}{c|}{-}          & 44.2       & 36.5       & 15.4       & 3.8        & -          \\
\multicolumn{1}{l|}{}                                     & \textbf{Student}              & 3.7        & \multicolumn{1}{c|}{4.0}          & 3.8        & \multicolumn{1}{c|}{4.0}          & 4.3               & 4.0                 & 8.3        & 50.0       & 41.7       & -          & \multicolumn{1}{c|}{-}          & -          & 91.7       & -          & 8.3        & \multicolumn{1}{c|}{-}          & 41.7       & 41.7       & 16.7       & -          & -          \\ \midrule
\multicolumn{1}{l|}{\multirow{3}{*}{\textbf{Experience}}} & \textbf{1-3 Years}            & 3.9        & \multicolumn{1}{c|}{4.0}          & 4.0        & \multicolumn{1}{c|}{4.0}          & 4.2               & 4.0                 & 28.1       & 39.1       & 26.6       & 6.3        & \multicolumn{1}{c|}{-}          & 26.6       & 48.4       & 18.8       & 6.3        & \multicolumn{1}{c|}{-}          & 45.3       & 32.8       & 18.8       & 3.1        & -          \\
\multicolumn{1}{l|}{}                                     & \textbf{4-6 Years}            & 4.4        & \multicolumn{1}{c|}{5.0}          & 4.4        & \multicolumn{1}{c|}{5.0}          & 4.4               & 5.0                 & 57.5       & 22.5       & 17.5       & 2.5        & \multicolumn{1}{c|}{-}          & 55.0       & 25.0       & 20.0       & -          & \multicolumn{1}{c|}{-}          & 55.0       & 30.0       & 15.0       & -          & -          \\
\multicolumn{1}{l|}{}                                     & \textbf{\textgreater 6 Years} & 4.6        & \multicolumn{1}{c|}{5.0}          & 4.5        & \multicolumn{1}{c|}{4.5}          & 4.6               & 5.0                 & 58.3       & 41.7       & -          & -          & \multicolumn{1}{c|}{-}          & 50.0       & 50.0       & -          & -          & \multicolumn{1}{c|}{-}          & 58.3       & 41.7       & -          & -          & -          \\ \bottomrule
\end{tabular}
}
\end{table}

\noindent\textbf{Likelihood of Using and Recommending \textit{AUTOCOMBAT}.} We assessed the likelihood of participants recommending \textit{AUTOCOMBAT} (Figure \ref{fig:recommendation}). The majority (84.5\%) expressed willingness to use or recommend the tool, with 37.9\% rating their likelihood as \textit{“Very Likely”} and 46.6\% as \textit{“Somewhat Likely.”} Neutral or negative responses were minimal, indicating strong user satisfaction and broad acceptance of the tool. 
Participants also suggested areas for improvement, with a primary focus on usability, feature enhancements, and code quality. Key recommendations included automating installation, incorporating a human-in-the-loop mechanism, and extending compatibility beyond Chrome. Additional suggestions involved verifying comment accuracy, integrating multiple top-rated solutions, and adapting the tool for other platforms. To further enhance answer quality, participants recommended adding executable examples, ensuring precise code generation, and highlighting key text segments. Nevertheless, the majority expressed satisfaction with \textit{AUTOCOMBAT}, with several participants stating \textit{“No suggestions”} or \textit{“Great as it is”}.


\begin{findingbox}
\textbf{RQ\textsubscript{3} Summary}:
Participants found \textit{AUTOCOMBAT} easy to install, effective in reducing editing effort, and broadly useful across professions and experience levels. A majority (84.5\%) were willing to use or recommend the tool. Suggestions focused on automating installation, expanding platform support, and enhancing code generation. Overall, feedback confirms \textit{AUTOCOMBAT}'s high usability and broad acceptance with minor improvement areas.
\end{findingbox}

\section{Discussion}
\label{sec:discussion}


\subsection{Key Findings}

\noindent\textbf{Effect of comment volume on LLM performance.}
Our experiments reveal two distinct effects of comment volume on LLM behavior. First, LLMs identify improvement concerns most effectively when answers have a moderate number of comments (around 2-5). In this range, feedback provides clear and relevant signals (real improvement requests) without introducing much noise (general talk), allowing models like DeepSeek to maintain a strong balance between precision and recall.  Second, when answers contain many comments ($\geq$6), the quality of the refined answers declines. Long discussion threads often blend useful insights with off-topic or redundant remarks, making it difficult for LLMs to distinguish genuine improvement suggestions from noise.

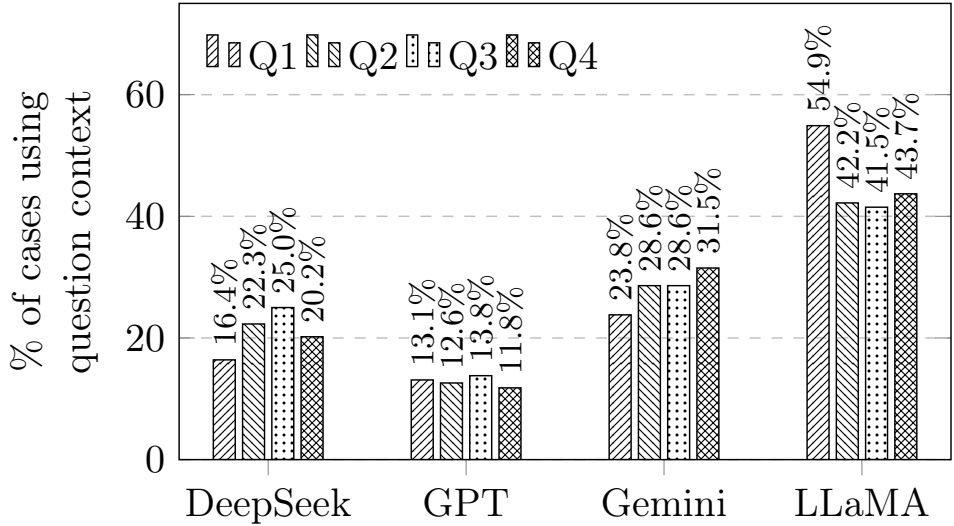
\begin{figure}[ht]
\centering
\resizebox{0.8\textwidth}{!}{%
\begin{tikzpicture}
\begin{axis}[
    ybar,
    bar width=6pt,
    width=9cm,
    height=6cm,
    enlarge x limits=0.15,
    ylabel={\% of cases using \\ question context},
    ylabel style={align=center},
    symbolic x coords={DeepSeek,GPT,Gemini,LLaMA},
    xtick=data,
    xtick pos=bottom,
    ymin=0, ymax=75,
    legend style={at={(0.02,0.95)}, anchor=north west, legend columns=4, draw=none}, 
    ymajorgrids=true,
    grid style=dashed,
    nodes near coords style={
    rotate=90, anchor=west, font=\footnotesize, text=black},
    nodes near coords={\pgfmathprintnumber[fixed,precision=1,fixed zerofill]{\pgfplotspointmeta}\%}
]

\addplot+[ybar,fill=white,draw=black,pattern=north east lines] 
coordinates {(DeepSeek,16.4) (GPT,13.1) (Gemini,23.8) (LLaMA,54.9)};

\addplot+[ybar,fill=white,draw=black,pattern=north west lines] 
coordinates {(DeepSeek,22.3) (GPT,12.6) (Gemini,28.6) (LLaMA,42.2)};

\addplot+[ybar,fill=white,draw=black,pattern=dots] 
coordinates {(DeepSeek,25.0) (GPT,13.8) (Gemini,28.6) (LLaMA,41.5)};

\addplot+[ybar,fill=white,draw=black,pattern=crosshatch] 
coordinates {(DeepSeek,20.2) (GPT,11.8) (Gemini,31.5) (LLaMA,43.7)};

\legend{Q1,Q2,Q3,Q4}
\end{axis}
\end{tikzpicture}
}
\caption{Question context usage by LLMs across Quartiles.}
\label{fig:q_context}
\end{figure}

\noindent\textbf{Balanced context use leads to more reliable refinements.} Our findings (Fig.~\ref{fig:q_context}) reveal clear differences in how LLMs rely on the corresponding question when refining SO answers. DeepSeek and Gemini use question context in 16-32\% of cases, increasing reliance in Q2-Q3 where answers have two to five comments. This suggests that both models rely mainly on comments but consult the question when feedback becomes moderately complex or ambiguous. In Q4, as comment threads become noisier, Gemini further increases its context use (31.5\%) while DeepSeek moderates it (20.2\%), reflecting adaptive grounding behavior.

GPT relies least on the question (11-14\%) across all quartiles, indicating a comment-driven approach that captures nearly all feedback but often over-flags irrelevant concerns. In contrast, LLaMA depends heavily on the question (41-55\%), especially in Q1 (54.9\%), revealing difficulty interpreting sparse feedback and a tendency to over-rely on context. This aligns with its lower precision and semantic alignment observed in \textit{RQ1} and \textit{RQ2}.

Consider a \texttt{SO} answer \cite{stackoverflow14960} where the user wanted to disable \texttt{Alt+F4} in a C\# WinForms dialog to prevent users from closing it. The code snippet in the original answer was:

\begin{lstlisting}[language=C]
private void Form1_FormClosing(object sender, FormClosingEventArgs e) {
    e.Cancel = true;
}
\end{lstlisting}

Comments pointed out that this blocks \emph{all} close attempts, including system shutdowns, and suggested instead:

\begin{lstlisting}[language=C]
e.Cancel = (e.Reason == CloseReason.UserClosing);
\end{lstlisting}

Here, the comments alone propose a fix but do not explain why it matters. The model needed to consult the original question text (\emph{“prevent the user from closing the form”}) to understand that only user-triggered closes (like \texttt{Alt+F4}) should be blocked, while system closes must still be allowed. Models that underuse question context (e.g., GPT) risk overgeneralizing this feedback, while models that apply balanced context use (e.g., DeepSeek) correctly limit the change to user-close events only. 

In contrast, consider another \texttt{SO} answer \cite{stackoverflow1309268} where the user wanted to flip an image horizontally or vertically using CSS for cross-browser support. The code snippet in the original answer was:

\begin{lstlisting}
.flip-horizontal {
    transform: scale(-1, 1);
    -moz-transform: scale(-1, 1);
    -webkit-transform: scale(-1, 1);
    filter: fliph; /*IE*/
}
.flip-vertical {
    transform: scale(1, -1);
    -moz-transform: scale(1, -1);
    -webkit-transform: scale(1, -1);
    filter: flipv; /*IE*/
}
\end{lstlisting}

Comments pointed out that Opera-specific CSS transforms were missing and suggested adding them:

\begin{lstlisting}
-o-transform: scale(-1, 0);
-o-transform: scale(0, -1);
\end{lstlisting}
\smallskip
Here, the comments themselves fully specified the required fix (add Opera-specific prefixes), and DeepSeek directly incorporated them into the answer. Since this refinement did not require understanding the original question’s intent, only appending missing browser-prefixed properties, the model correctly did not consult the question text. These demonstrate how selective context grounding improves alignment with the asker’s intent and leads to more reliable refinements.


\noindent\textbf{\textit{AUTOCOMBAT} produces revisions that closely resemble human edits.} 
Comparing \textit{AUTOCOMBAT}’s outputs with human-revised answers showed comparable quality, with mean BLEURT 0.72 and COMET 0.79. By balancing comment-based feedback with relevant question context, the system replicated typical developer edits such as clarifying explanations, updating code, and improving structure. Its strength lies in synthesizing multiple concerns into cohesive revisions, mirroring how human editors integrate diverse feedback into balanced, context-aware improvements.


\noindent\textbf{Preserving technical intent while integrating community feedback.} The intent preservation analysis reveals that high surface similarity between LLM-enhanced and human-revised answers does not necessarily imply alignment in technical intent. LLMs that balance precision and recall in identifying improvement-related concerns, particularly DeepSeek, are more successful at refining answers without altering the original problem scope or assumptions. In contrast, models that emphasize recall tend to introduce partial intent drift by expanding explanations or incorporating changes beyond those applied by human editors. An interesting observation is that answers associated with longer comment threads often exhibit higher intent preservation, suggesting that well-articulated feedback can help anchor refinements when actionable concerns are appropriately filtered. Nevertheless, the persistence of partial intent deviations highlights the challenge of integrating diverse user perspectives while maintaining fidelity to the original technical goal, reinforcing intent preservation as a critical dimension of quality in automated answer enhancement.



\noindent\textbf{High adoption potential for real-world impact.} Our user study showed that over 84\% of developers are willing to adopt and recommend \textit{AUTOCOMBAT}. Participants found the tool easy to use, efficient, and effective in enhancing SO answers, demonstrating its practical value for real-world applications.

\noindent\textbf{Users valued clarity and coverage over technical perfection.} From the user study, developers consistently rated \textit{AUTOCOMBAT} highest on \textit{clarity} (readability and structure) and \textit{coverage} (how well concerns were addressed), even more than raw technical precision. This indicates that when evaluating answer quality, practitioners care about how understandable and comprehensive an answer feels, not just whether the code is technically correct. The correlation here is striking: it aligns with prior findings on SO that well-explained answers are more trusted than brief, technically dense ones \cite{kabir2024stack}.


\subsection{Implications}
\begin{figure}[!htbp]
    \centering
    \begin{subfigure}{\columnwidth}
        \centering
        \includegraphics[width=4in]{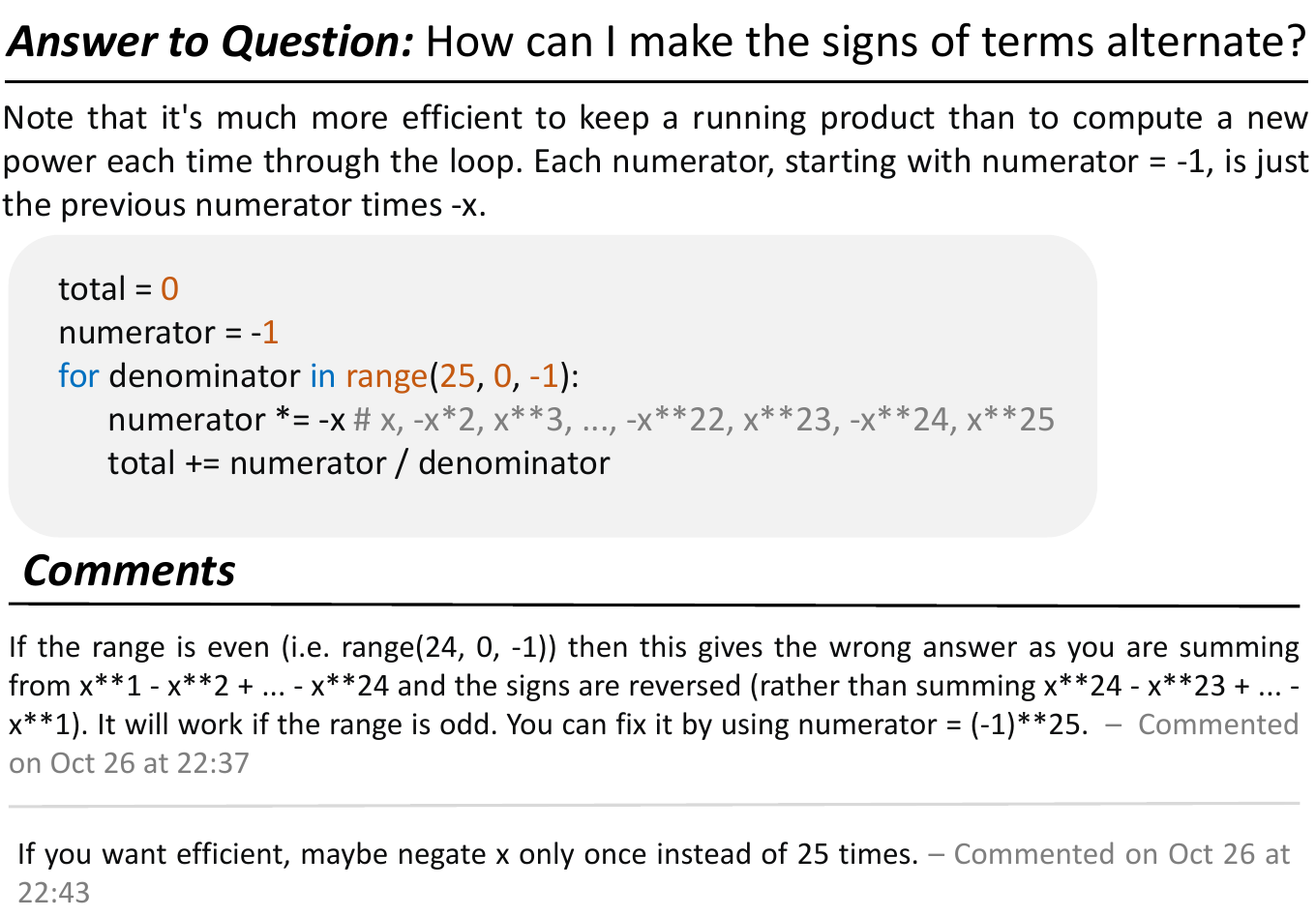}
        \caption{Initial answer.}
        \label{fig:motex_original_version}
    \end{subfigure}
    
    \begin{subfigure}{\columnwidth}
        \centering
        \includegraphics[width=4.2in]{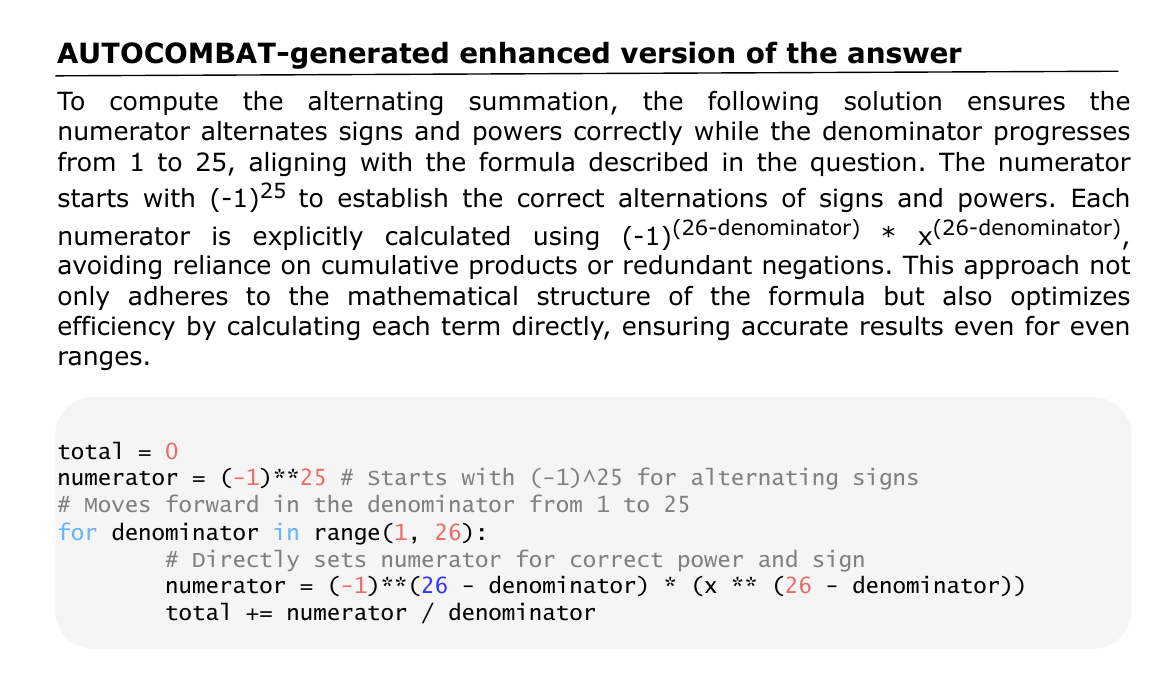}
        \caption{Enhanced answer produced by \textit{AUTOCOMBAT}.}
        \label{fig:motex_enchaned_version}
    \end{subfigure}

    \caption{An example \cite{motx} shows an SO answer with multiple unaddressed user concerns and the improved version automatically generated by \textit{AUTOCOMBAT}.}
    \label{fig:overall_mot_ex}
\end{figure}

Our findings show that feedback-driven refinement can make programming knowledge more accurate, reliable, and sustainable when guided by human-aligned LLMs. Fig.~\ref{fig:motex_original_version} illustrates this potential: an original SO answer attempted to optimize a mathematical series computation but left critical issues unresolved, including a sign error and inefficient negation.
After analyzing the associated user comments, \textit{AUTOCOMBAT} synthesized the feedback and generated an improved version (Fig.~\ref{fig:motex_enchaned_version}), automatically correcting both logic and clarity. This example highlights how structured feedback integration enables LLMs to perform contextually faithful refinements that previously required manual effort.

Beyond technical improvements, this capability reshapes how knowledge evolves within developer communities. For \textbf{Developers}, \textit{AUTOCOMBAT} serves as a collaborative assistant that consolidates scattered feedback into coherent updates, reducing the effort of reconciling conflicting suggestions and helping keep answers relevant and up to date. For \textbf{Researchers}, the tool and the \textit{ReSOlve} benchmark offer a foundation to study human-LLM collaboration, adaptive feedback modeling, and explainable refinement. For \textbf{Platform Owners}, such as SO, \textit{AUTOCOMBAT} provides a scalable way to preserve content quality, reduce moderation workload, and sustain engagement through continuous feedback integration.

\section{Threats to Validity}
\label{sec:threats}

\noindent\textbf{External validity} concerns the generalizability of our findings. Our results may not generalize to all the answers on SO. We analyzed four widely used programming languages: Python, Java, C\#, and JavaScript to capture diverse paradigms and used quartile-based sampling to include answers with varying comment volume.
To strengthen the validity of our conclusions, we sampled 1,000 answers (reduced to 790 after filtering), which is statistically significant with a 95\% confidence level and 5\% error margin.
We evaluated four state-of-the-art LLMs (DeepSeek, GPT, Gemini, LLaMA) to reduce model-specific bias, which showed consistent trends. 

\smallskip
\noindent\textbf{Internal validity} relates to potential process bias. Manual annotation may introduce subjectivity bias, but two annotators, calibration sessions, and high agreement ($\kappa > 0.9$) mitigated this bias. 
For the user study, we recognized that the snowball approach can introduce selection bias by overrepresenting participants within similar professional networks. To mitigate this, we complemented it with open recruitment to broaden participant diversity and applied anonymity to encourage honest and unbiased responses. Individual bias was further reduced by involving 58 professionals from varied domains and experience levels, though the sample may not fully represent the entire developer population.

\smallskip
\noindent\textbf{Construct validity} assesses whether our measures capture answer improvement quality. The identification of improvement-related comments was evaluated using standard metrics (accuracy, precision, recall, F1, specificity, MCC), and refined answers were assessed with syntactic and semantic metrics (e.g., ROUGE, BLEU, METEOR, TER, BLEURT, COMET, BERTScore). Because these metrics may not fully capture readability or practical usefulness, we conducted a user study evaluating \textit{coverage}, \textit{clarity}, and \textit{relevance} to reflect developers’ perceptions. To further verify functional soundness, we manually reviewed \textit{AUTOCOMBAT} refinements and found that 88.10\% fully preserved the original intent, while 10.63\% exhibited partial intent preservation. These results indicate that automated metrics are well aligned with human-perceived quality, and that targeted human validation can further enhance confidence in behavior-critical refinements.

\section{Related Work}
\label{sec:related-work}

This section reviews prior work on LLMs in SE, studies on SO answer quality, and efforts to improve those answers.

\smallskip
\noindent\textbf{LLMs in SE.}
LLMs are increasingly applied to code generation, repair, documentation, and testing. Chen et al. \cite{chen2021evaluating} introduced \textit{Codex} and the HumanEval benchmark, showing that transformer-based models can generate functionally correct code from natural-language prompts. Beyond general-purpose models, SE-specific pretraining has proven effective: CodeT5 unifies code understanding and generation tasks under an encoder–decoder framework and improves performance across multiple benchmarks \cite{wang2021codet5}. For documentation-related tasks, transformer-based approaches advance code summarization quality \cite{ahmad2020transformer}. LLMs are also being leveraged for bug fixing and automated program repair (APR). Two studies showed that large pre-trained models can generate competitive patches and that APR workflows can be redesigned around LLM capabilities: Xia et al.demonstrated strong LLM-based APR with careful prompting and context \cite{xia2023automated}, while Fan et al. examined repairing LLM-generated buggy code and how APR tools/LLMs complement each other \cite{fan2023automated}. Human-in-the-loop workflows highlight how developers actually use these tools. Vaithilingam et al. conducted a controlled study of GitHub Copilot with professional programmers, reporting productivity gains alongside the need for oversight and task decomposition strategies \cite{vaithilingam2022expectation}. Mastropaolo et al. provided an empirical analysis of Copilot’s outputs against tests at scale, documenting quality and correctness considerations developers must validate \cite{mastropaolo2023robustness}. Finally, community benchmarks such as APPS \cite{hendrycks2021measuring} and open models like Code Llama \cite{roziere2023code} have further standardized evaluation and broadened access to capable code LLMs. Peng et al. \cite{peng2023impact} and Barke et al. \cite{barke2023grounded} found that developers often edit LLM-generated code for correctness and efficiency, emphasizing the need for feedback-grounded refinement mechanisms.   

The above studies primarily focused on generation accuracy, repair effectiveness, and productivity gains, but the human-aligned improvement of code based on developer feedback and real-world concerns remains understudied.

\smallskip
\noindent\textbf{SO Answer Quality.}
Research on SO answer quality has examined both prediction and improvement. Shah and Pomerantz \cite{shah2010evaluating} evaluated quality using clarity and completeness, while Bloom et al. \cite{blooma2010selection} emphasized content value over engagement. Later works combined content, code, and user features for best-answer prediction \cite{ercan2015automatic, zheng2017best, roy2024predicting}. Others developed tools for detecting or improving low-quality posts, such as SOPI \cite{tavakoli2016improving} and AnswerBot \cite{cai2019answerbot}. Comments play a crucial role in answer evolution: Zhang et al. \cite{zhang2019empirical} found that many answers become outdated, with 69\% of weak C/C++ snippets remaining unaddressed \cite{Zhang2022A}. Ragkhitwetsagul et al. \cite{ragkhitwetsagul2019toxic} and Wang et al. \cite{wang2018users} showed that while comments often trigger edits, most are minor. Baltes et al. \cite{baltes2018sotorrent, baltes2019sotorrent} and Diamantopoulos et al. \cite{diamantopoulos2019towards} linked edits to social incentives rather than content needs.
%

A recent study proposed SOUP \cite{Mai2024TowardsBA}, a framework for refining SO code snippets using user feedback. For answers involving a single comment and a single code snippet, SOUP demonstrates promising performance in improving code snippets. However, SOUP has several notable limitations: it cannot enhance textual explanations, cannot handle multiple code snippets, processes only single-comment feedback, and overlooks the broader question context. As a result, it often produces syntactically valid but semantically inconsistent updates.
%
Building on these findings, we propose \textit{AUTOCOMBAT}, a feedback-aware LLM-powered tool that integrates multiple comment concerns and leverages question context to refine both text and code, delivering holistic and context-sensitive improvements to SO answers.

\section{Conclusion and Future Work}
\label{sec:conclusion}

LLMs have demonstrated impressive capabilities in code generation, documentation, and optimization. However, their potential to improve existing programming answers in a manner aligned with human revision practices has remained underexplored. In this study, we introduce \textit{AUTOCOMBAT}, a tool that leverages LLMs to enhance SO answers in a human-aligned manner through structured integration of community feedback and contextual reasoning. Our findings demonstrate that LLMs can effectively identify improvement-related concerns expressed in user comments, with DeepSeek offering the most balanced and reliable performance. Building on this capability, \textit{AUTOCOMBAT} refines answers by systematically addressing actionable feedback while preserving the original technical intent. Consequently, it produces revisions that are comparable to human edits, consistently outperforms SOUP, and is well received by practitioners, with 84.5\% of participants reporting that they would use or recommend the tool.

Overall, this study reframes programming answer improvement as a process of human–LLM collaboration, where LLMs act as responsive assistants that synthesize distributed community feedback into coherent, intent-preserving enhancements. By operationalizing this paradigm at scale, \textit{AUTOCOMBAT} demonstrates how LLMs can contribute to the long-term maintenance, reliability, and trustworthiness of community-driven programming knowledge.

Looking ahead, future work will extend this approach to support multi-turn feedback and real-time integration with other Q\&A platforms (e.g., GitHub Discussions) as well as domain-specific forums, enabling broader and more sustained improvement of community-driven programming knowledge.

\section{Acknowledgments}

This research is supported in part by the Natural Sciences and Engineering Research Council of Canada (NSERC) Discovery Grants program, the Canada Foundation for Innovation's John R. Evans Leaders Fund (CFI-JELF), and by the industry-stream NSERC CREATE in Software Analytics Research (SOAR).

\bibliographystyle{ieeetr}
\bibliography{bibliography}












\end{document}